\documentclass[prd,nofootinbib,twocolumn]{revtex4}
\usepackage{amssymb}
\usepackage{amsmath}
\usepackage{natbib}
\usepackage{graphicx}

\usepackage{url}
\def\mbf#1{\ensuremath{\mathchoice{\mbox{\boldmath$\displaystyle#1$}}
{\mbox{\boldmath$\textstyle#1$}}
{\mbox{\boldmath$\scriptstyle#1$}}
{\mbox{\boldmath$\scriptscriptstyle#1$}}}}

\def\mb{\mbox}

\def\idm{\mathbb{I}}
\def\const{\mathrm{const}}

\def\Tr{\textrm{Tr}}

\def\fkdot{f^{(k)}}
\def\fdot{{\dot{f}}}
\def\tauref{\tau_\mathrm{ref}}

\def\F{\mathcal{F}}

\def\N{\mathcal{N}}
\def\A{\mathcal{A}}
\def\C{\mathcal{C}}
\def\O{\mathcal{O}}

\def\R{\mathcal{R}}
\def\L{\mathcal{L}}
\def\G{\mathcal{G}}

\def\S{\mathcal{S}}
\def\Sinv{\S^{-1}}

\def\parm{\theta}

\def\M{\mathcal{M}}
\def\Minv{\M}

\def\X{{\mathrm{X}}}
\def\Y{{\mathrm{Y}}}
\def\Z{{\mathrm{Z}}}

\def\trans{{^\mathrm{T}\!}}

\def\mhat{{\widehat{m}}}
\def\mtilde{{\widetilde{m}}}

\def\vn{{\vec{n}}}
\def\vr{{\vec{r}}}

\def\av#1{{\langle #1 \rangle}}
\def\avS#1{{\av{#1}_{\!S}}}

\def\di{{\partial_i}}
\def\dj{{\partial_j}}

\def\min{{\mathrm{min}}}
\def\max{{\mathrm{max}}}

\def\dopVec#1{#1}
\def\detVec#1{\mbf{#1}}
\def\ampVec#1{#1}

\def\Doppler{\lambda}
\def\vDoppler{{\dopVec{\Doppler}}}

\def\Intrins{\omega}
\def\vA{{\ampVec{\A}}}

\def\sig{{\mathrm{s}}}

\def\gt{\widetilde{g}}
\def\gh{\widehat{g}}

\def\dth{\Delta\theta}

\def\gbar{\overline{g}}
\def\mbar{\overline{m}}

\def\Ndet{\mathcal{N}}

\def\orb{{\mathrm{orb}}}

\def\spin{{\mathrm{spin}}}

\def\qu{{\mathrm{qu}}}

\def\relError{\varepsilon}
\def\cosi{{\cos\iota}}

\def\dOm{{\Delta\widehat{\Omega}}}
\def\uncertF{\sigma_\F}

\def\rad{\textrm{rad}}

\def\vsep{0.1cm}

\begin{document}

\title{The search for continuous gravitational waves: metric of
  the multi-detector $\F$-statistic}  
\author{Reinhard Prix}
\affiliation{Max-Planck-Institut f\"ur Gravitationsphysik,
Albert-Einstein-Institut, Am M\"uhlenberg 1,
14476 Golm, Germany}

\begin{abstract}
We develop a general formalism for the parameter-space metric of the
multi-detector $\F$-statistic, which is a matched-filtering detection
statistic for continuous gravitational waves. 
We find that there exists a whole \emph{family} of $\F$-statistic metrics,
parametrized by the (unknown) amplitude parameters of the
gravitational wave. 
The multi-detector metric is shown to be expressible in terms of
noise-weighted \emph{averages} of single-detector contributions, which
implies that the number of templates required to cover the parameter
space does \emph{not} scale with the number of detectors.    
Contrary to using a longer observation time, combining detectors of
similar sensitivity is therefore the computationally
cheapest way to improve the sensitivity of coherent wide-parameter
searches for continuous gravitational waves.  

We explicitly compute the $\F$-statistic metric family for signals
from isolated spinning neutron stars, and we numerically evaluate the
quality of different metric approximations in a Monte-Carlo study. 
The metric predictions are tested against the measured mismatches and
we identify regimes in which the local metric is no longer a good
description of the parameter-space structure. 
\end{abstract}

\preprint{AEI-2006-043}

\maketitle

\section{Introduction}

Continuous gravitational waves (GWs), which would be emitted, for
example, by spinning non-axisymmetric neutron stars, or by solar-mass
binary systems, are generally expected to be so weak that they will be
buried several orders of magnitude below the noise of even the most
sensitive detectors.   
The detection of such signals therefore requires the exact knowledge
of their waveform, in order to be able to coherently correlate the
data with the expected signal by \emph{matched filtering}. 

In a wide-parameter search for unknown sources, however, we typically
only know the family of possible waveforms (or an approximation
thereof), parametrized by unknown signal parameters (such as the
frequency or sky position of the source). 
The corresponding parameter space needs to be covered by a
finite number of ``templates'' for which a search will be performed.
These templates must be placed densely enough, so that for any
possible signal, no more than a certain fraction of the
signal-to-noise ratio (SNR) is lost at the closest template. On the
other hand, coherently correlating the data with every templates is
computationally expensive and increases the number of statistical
false-alarm candidates. Therefore an \emph{optimal} covering 
is desirable, which minimizes the number of templates but guarantees
the required ``minimal match''.

In order to solve this \emph{covering problem}, it is essential to
understand the underlying parameter-space structure. 
Most studies on the construction of optimal template banks have
been performed in the context of binary-inspiral searches.
It was realized early on that a geometric approach to this problem is
the most natural, in particular the introduction of a \emph{metric} on
the parameter space by \citet{bala96:_gravit_binaries_metric} and
\citet{owen96:_search_templates}, building on the earlier concept of
the  ``fitting factor'' introduced by \citet{apostolatos95:_search_templates}.   
Note, however, that this definition of the metric differs subtly from
the ``canonical'' definition used in the present work (and also in
\cite{krolak04:_optim_lisa}), which is derived directly from the
detection statistic (see appendix~\ref{sec:altern-proj-onto} for more
details).  
The canonical definition of the metric assigns the relative loss of
SNR due to an offset in signal parameters as an invariant ``distance''
measure, which can locally be expressed  as a metric tensor. 
The metric is closely related to the well-known concept of the
``Fisher information matrix'',  which quantifies the statistical
errors in the parameter estimation of signals:  the (canonical) metric
is identical to the \emph{normalized} Fisher matrix, even though it
describes conceptually rather different aspects of the detection
statistic.   

A somewhat related question is the \emph{global} parameter-space
structure, which was studied in \cite{prix05:_circles_sky} for the
case of isolated neutron-star signals. This study found that the global
structure (the ``circles in the sky'') deviates significantly from the
local metric picture. 
The global structure is relevant, for example, for deciding whether
different detection candidates are consistent with the same signal,
i.e. whether they are ``coincident candidates''. Obviously, the metric
description is the local approximation to this global parameter-space
structure. 

In this paper we consider gravitational-wave signals that are nearly
monochromatic and sinusoidal in the frame of the GW source, and which
are of long duration (i.e. typically longer than the observation time $T$).
This class of signals is usually referred to as ``continuous waves'',
and the prime examples are GWs from non-axisymmetric spinning neutron
stars (e.g. see \cite{prix06:_cw_review} for a review)
and stellar-mass binary systems in the LISA frequency band (e.g. see
\cite{krolak04:_optim_lisa}).  

The phase of the signal received at the detector is 
Doppler modulated by the rotation and orbital motion of the detector.
The observed phase therefore depends not only on the intrinsic
frequency evolution of the signal, but also on its sky position. 
In addition to the phase modulation, there is a time-varying
amplitude modulation, due to the rotating antenna pattern of the
detector. This amplitude modulation depends on the polarization angle
$\psi$ and the polarization amplitudes $A_+$ and $A_\times$ of the GW. 
However, as shown by \citet{jks98:_data}, these unknown parameters
(together with the initial phase $\phi_0$), can be eliminated by
analytically maximizing the detection statistic.
The resulting reduced parameter space includes only the parameters
affecting the time evolution of the signal phase, which we will refer
to as the ``Doppler parameters''.
This amplitude-maximized detection statistic is generally known as the
``$\F$-statistic'', which has been used in several searches for
continuous GWs from spinning neutron stars 
(e.g. \cite{lsc04:_psr_j1939,lsc06:_coher_scorp_x,2005CQGra..22S1243A}).   
After two earlier (partly successful) attempts
to generalize the $\F$-statistic to a coherent network of detectors
\cite{jks98:_data,krolak04:_optim_lisa}, this problem was fully solved
more recently by \citet{cutler05:_gen_fstat}.  

Somewhat surprisingly, however, there has not been much work on
the metric of the $\F$-statistic, neither in the single- nor the
multi-detector case: the single-detector $\F$-statistic metric was
derived on a formal level by \citet{krolak04:_optim_lisa}, but was not
evaluated explicitly or studied further.   
A single-detector $\F$-statistic metric was used 
(without giving any details) in \cite{cornish05:_detec_lisa}
to numerically estimate the number of templates in galactic-binary 
searches with LISA. 
An earlier study by \citet{brady98:_search_ligo_periodic}
of the metric for isolated neutron-star signals introduced a metric
approximation based only on the phase modulation of the signal and
neglecting the amplitude modulation. 
We will refer to this approximation as the ``phase metric''. 
This metric has a simpler structure than the full $\F$-metric, and it
can be computed analytically \cite{jones05:_ptole_metric} if one
assumes a circular orbital motion. This is the only type of
continuous GW metric that is currently implemented in
LAL/LALApps~\cite{lalapps}.  

As we will see in this study, the amplitude modulation cannot always
be neglected, but ``on average'' the phase metric seems to be a good
approximation, and its quality improves with longer observation times
and with the number of detectors. 
With the recent multi-detector generalization of the 
$\F$-statistic formalism~\cite{cutler05:_gen_fstat} and its subsequent
implementation into LAL/LALApps, the need to understand the
\emph{multi-detector} $\F$-statistic metric has become more urgent.  
The most important question in this context is whether the metric
resolution increases with the number of detectors, i.e. whether a
denser covering of the parameter space is required, which would 
increase the computational cost.

The main result of this work is to show that the metric resolution does
\emph{not} scale with the number of detectors. 
Therefore, sensitivity can be gained at the cost of only a linear
increase in the required computing power (as the signal has to be
correlated with the data stream from each detector). 
This has to be contrasted with the (at least) $\O(T^5)$ scaling
\eqref{eq:113} of the number of templates with observation time $T$,
in the case isolated neutron-star signals with one spindown. 
  
In order to improve the sensitivity of a coherent search for
continuous gravitational waves, increasing the number of
similar-sensitivity detectors is therefore computationally much
cheaper than to increase the observation time.

The plan of this paper is as follows:
in Sect.~\ref{sec:multi-detector-f} we introduce the formalism and
notation of the multi-detector $\F$-statistic, following
\cite{cutler05:_gen_fstat} and \cite{krolak04:_optim_lisa}.
In Sect.~\ref{sec:f-metric} we derive the $\F$-statistic
metric family for high-frequency signals (relevant for ground-based
detector networks).  
We compute the extremal range of this metric family, its average
metric, and the long-duration limit, in which the $\F$-metric family
reduces to a simple ``orbital metric''. 
In Sect.~\ref{sec:appl-isol-puls} we apply this framework to the
special case of GWs from isolated spinning neutron stars, and we
evaluate the quality of the metric predictions (and different
approximations) by comparing them against measured mismatches in a
Monte-Carlo study. 
The main results are summarized in Sect.~\ref{sec:conclusions}. 
Appendix~\ref{sec:altern-proj-onto} presents an alternative, more
elegant derivation of the $\F$-statistic metric, and
Appendix~\ref{sec:keep-ampl-funct} gives the general expressions
for the $\F$-metric, which would be valid also for low-frequency
signals relevant for LISA.

\section{The multi-detector $\F$-statistic}
\label{sec:multi-detector-f}

\subsection{General definitions}
\label{sec:general-case}

In this section we introduce the formalism and notation of 
the $\F$-statistic, a matched-filtering detection statistic for
continuous gravitational waves, which was first introduced by
\citet{jks98:_data}, and subsequently generalized to the
multi-detector case by \citet{cutler05:_gen_fstat}.
As shown in \cite{jks98:_data}, the dimensionless strain signal
$s^\X(t)$ of a continuous gravitational wave at detector $\X$ can be
represented in the form  
\begin{equation}
  \label{eq:133}
  s^\X(t) = \sum_{\mu=1}^4 \A^\mu \, h^\X_\mu(t)\,,
\end{equation}
in terms of four signal-amplitudes $\A^\mu$, which are
independent of the detector $\X$, and the detector-dependent basis
waveforms $h_\mu^\X(t)$. 
The four amplitudes $A^\mu$ can be expressed in terms of two
polarization amplitudes $A_+$, $A_\times$, the initial phase $\phi_0$
in the solar-system barycenter (SSB) at a reference time $\tauref$,
and the polarization angle $\psi$ of the wave frame with respect to
the equatorial coordinate system, namely   
\begin{equation}
  \label{eq:60}
  \begin{array}{r c l}
    \A^1 &=&\;\;A_+\,\cos\phi_0\,\cos 2\psi - A_\times\,\sin\phi_0\,\sin 2\psi\,,\\[\vsep]
    \A^2 &=&\;\;A_+\,\cos\phi_0\,\sin 2\psi + A_\times\,\sin\phi_0\,\cos 2\psi\,,\\[\vsep]
    \A^3 &=&-A_+\,\sin\phi_0\,\cos 2\psi - A_\times\,\cos\phi_0\,\sin 2\psi\,,\\[\vsep]
    \A^4 &=&-A_+\,\sin\phi_0\,\sin 2\psi + A_\times\,\cos\phi_0\,\cos 2\psi\,.\\ 
    \end{array}
\end{equation}
We can further relate the two polarization amplitudes $A_+$ and $A_\times$ to
the overall amplitude $h_0$ and the inclination angle $\iota$ of the
quadrupole rotation axis with respect to the line of sight, namely
\begin{equation}
  \label{eq:62}
  A_+ = \frac{1}{2} h_0 \, \left( 1 + \cos^2 \iota \right)\,,\quad
  A_\times = h_0 \, \cosi\,.
\end{equation}
The four basis waveforms $h_\mu^\X(t)$ can be written as
\begin{equation}
  \label{eq:135}
  \begin{array}{l}
    h_1^\X(t) = a^\X(t) \cos\phi^\X(t)\,,\;
    h_2^\X(t) = b^\X(t) \cos\phi^\X(t)\,,\\[\vsep]
    h_3^\X(t) = a^\X(t) \sin\phi^\X(t)\,,\;
    h_4^\X(t) = b^\X(t) \sin\phi^\X(t)\,,\\
  \end{array}
\end{equation}
where $a^\X(t)$ and $b^\X(t)$ are the antenna-pattern functions
(see Eqs.(12,13) of \cite{jks98:_data}), and $\phi^\X(t)$ is the
signal phase at the detector $\X$.  
The antenna-pattern functions $a^\X(t),\,b^\X(t)$ depend on the
sky position $\vn$ of the GW source, and on the location and 
orientation of the detector $\X$. The phase $\phi^\X(t)$ also depends
on the \emph{intrinsic} phase parameters, $\Intrins$ say, of the signal. 
In the case of continuous waves from isolated neutron stars, $\Intrins$
would only consist of the $s+1$ spin parameters, i.e.
$\Intrins = \{f^{(k)}\}_{k=0}^{s}$, where $f^{(k)}$ is the k-th
time-derivative of the intrinsic signal frequency in the SSB (see
Sect.~\ref{sec:appl-isol-puls}). 
In the case of signals from spinning neutron stars in a binary
system, $\Intrins$ would also contain the binary orbital
parameters. We can summarize these dependencies as 
\begin{eqnarray}
  \label{eq:1}
  a^\X = a^\X(t; \vn)\,,\;\;
  b^\X = b^\X(t; \vn)\,,\;\;
  \phi^\X = \phi^\X(t; \vn, \Intrins)\,.
\end{eqnarray}
In the following we denote the set of ``Doppler parameters''
(i.e. the parameters affecting the time evolution of the phase)
by $\dopVec{\Doppler} \equiv \{\vn, \Intrins\}$, as opposed to the 
four ``amplitude parameters'' $\{\A\}^\mu = \A^\mu$.
Note that in the literature the Doppler parameters are sometimes
referred to as ``intrinsic'', and the amplitude parameters as
``extrinsic'', but we will not use this convention here. 

Using the multi-detector notation of
\cite{cutler05:_gen_fstat,krolak04:_optim_lisa}, 
we write vectors in ``detector-space'' in boldface,
i.e. $\{\detVec{s}\}^\X = s^\X$, and so the signal model
\eqref{eq:133} can be written as   
\begin{equation}
  \label{eq:102}
  \detVec{s}(t; \vA, \vDoppler) = \A^\mu\, \detVec{h}_\mu(t; \vDoppler)\,,
\end{equation}
with implicit summation over repeated amplitude indices, $\mu
\in\{1,2,3,4\}$. 
We assume the data $x^\X(t)$ of detector $\X$ contains 
a signal with parameters $\{\vA,\vDoppler\}$ in addition to
Gaussian stationary noise $n^\X(t)$, i.e.
\begin{equation}
  \label{eq:2}
  \detVec{x}(t) = \detVec{n}(t) + \detVec{s}(t; \vA, \vDoppler)\,.
\end{equation}
In general, the noise contributions $n^\X(t)$ of the different
detectors could be correlated (which might be relevant for the two
LIGO detectors in Hanford, or for LISA), and so we define the 
(double-sided) noise-density matrix $S^{\X\Y}$ as 
\begin{equation}
  \label{eq:71}
  S^{\X\Y}(f) = \int_{-\infty}^\infty \kappa^{\X\Y}(\tau) \, e^{-i 2\pi\,f
    \tau}\,d \tau\,,
\end{equation}
in terms of the correlation functions 
\begin{equation}
  \label{eq:75}
  \kappa^{\X\Y}(\tau) \equiv E\left[ n^\X(t + \tau)\, n^\Y(t) \right]\,.
\end{equation}
The corresponding multi-detector scalar product is defined (in analogy
to \cite{finn92:_detection}) as    
\begin{equation}
  \label{eq:95}
  \left(\detVec{x} | \detVec{y}\right) \equiv \int_{-\infty}^\infty 
  \widetilde{x}^\X(f)\, S^{-1}_{\X\Y}(f)\, \widetilde{y}^{\Y*}(f)\, d f\,,
\end{equation}
where $\widetilde{x}(f)$ denotes the Fourier-transformed of $x(t)$.
We use implicit summation over repeated detector indices, and
the inverse noise matrix is defined by
\mbox{$S^{-1}_{\X\Y}\,S^{\Y\Z} = \delta_\X^\Z$}. 
In the case of uncorrelated noise, where 
{$S^{\X\Y} = S^\X\,\delta^{\X\Y}$}, the scalar product simplifies to  
 \begin{equation}
  \label{eq:98}
  (\detVec{x}|\detVec{y}) = \sum_\X ( x^\X| y^\X ) \,,
\end{equation}
in terms of the usual single-detector scalar product
\begin{equation}
  \label{eq:99}
  (x^\X | y^\X ) \equiv \int_{-\infty}^\infty {\widetilde{x}^\X(f)\,
    \widetilde{y}^{\X*}(f) \over S^{X}(f) } \, d f\,.
\end{equation}
With the definition \eqref{eq:95} of the multi-detector scalar
product, the likelihood function for Gaussian stationary noise
$\detVec{n}(t)$ can be written as
\begin{equation}
  \label{eq:96}
  P\left( \detVec{n}(t) | S^{\X\Y} \right) = k \, 
  e^{-{1\over2}\left(\detVec{n}| \detVec{n}\right)}\,,
\end{equation}
where $k$ is a normalization constant.
Using \eqref{eq:2} we can express the likelihood of
observing data $\detVec{x}(t)$ in the presence of a signal 
$\detVec{s}(t; \vA, \vDoppler)$ as
\begin{equation}
  \label{eq:100}
  P( \detVec{x} | \vA, \vDoppler, S^{\X\Y}) = k\, e^{-{1\over 2}(\detVec{x}|\detVec{x})}\,
  e^{(\detVec{x}|\detVec{s}) - {1\over 2}(\detVec{s}|\detVec{s})} \,. 
\end{equation}
Using a standard frequentist approach, the optimal detection statistic
to decide between the hypothesis of a signal 
$s = \detVec{s}(\vA,\vDoppler)$ being present and no signal, $s=0$, is
given by the likelihood ratio $\Lambda(\detVec{x};\detVec{s})$, defined as
\begin{equation}
  \label{eq:26}
  \ln \Lambda(\detVec{x};\detVec{s}) \equiv \ln \frac{P(\detVec{x}|s)}{P(\detVec{x}|0)}
  = (\detVec{x}|\detVec{s}) - \frac{1}{2}(\detVec{s}|\detVec{s})\,.
\end{equation}
Substituting the signal model \eqref{eq:133}, the log-likelihood ratio
is found as  
\begin{equation}
  \label{eq:6}
  \ln \Lambda(\detVec{x}; \vA, \vDoppler) = \A^\mu \,x_\mu 
  - {1\over2} \A^\mu \,\M_{\mu\nu}\,\A^\nu\,,
\end{equation}
where we defined   
\begin{eqnarray}
  \label{eq:7}
  x_\mu(\vDoppler) &\equiv& \left(\detVec{x} |\detVec{h}_\mu \right)\,,\\
  \M_{\mu\nu}(\vDoppler) &\equiv& \left( \detVec{h}_\mu | \detVec{h}_\nu\right)\,.
  \label{eq:25}
\end{eqnarray}
We see that the likelihood ratio \eqref{eq:6} can be maximized analytically
with respect to the unknown amplitudes $\A^\mu$, and the resulting
detection statistic for the Doppler parameters $\vDoppler$ is the
so-called \emph{$\F$-statistic}, namely 
\begin{equation}
  \label{eq:8}
  2\F(\detVec{x};\vDoppler) \equiv x_\mu\,\Minv^{\mu\nu}\, x_\nu\,,
\end{equation}
where $\M^{\mu\nu} \equiv \{\M^{-1}\}^{\mu\nu}$, 
i.e.\ $\M_{\mu\alpha}\M^{\alpha\nu} = \delta_\mu^\nu$. 
Note that the four (multi-detector) wave-functions $\detVec{h}_\mu(t)$
form a basis of the signal space, and $\M$ is the associated metric,
which allows us to raise and lower amplitude indices. 
In particular, we could define a ``dual'' basis, 
$\detVec{h}^\mu(t) \equiv \M^{\mu\nu}\,\detVec{h}_\nu(t)$, 
satisfying $(\detVec{h}^\mu | \detVec{h}_{\nu}) = \delta^\mu_\nu$, and
the $\F$-statistic \eqref{eq:8} could then be written even more
compactly as $2\F(\vDoppler) = x^\mu\, x_\mu$.

\subsection{$\F$-statistic of perfectly matched signals}
\label{sec:f-stat-perf}

Let us consider the case where the target Doppler parameters
$\vDoppler$ are perfectly matched to the signal $\vDoppler_\sig$,
so the measured data would be 
$\detVec{x}(t) = \detVec{n}(t) + \detVec{s}(t; \vA, \vDoppler)$.
In this case, the projections \eqref{eq:7} are 
\begin{equation}
  \label{eq:10}
  x_\mu(\vA,\vDoppler) = n_\mu(\vDoppler) + s_\mu(\vA,\vDoppler)\,,
\end{equation}
where we defined
$n_\mu \equiv \left(\detVec{n} | \detVec{h}_\mu\right)$  
and
$s_\mu \equiv \left(\detVec{s}| \detVec{h}_\mu\right)$.
Assuming Gaussian stationary noise, one can show that
\begin{equation}
  \label{eq:11}
  E\left[ n_\mu \right] = 0\,,\quad\textrm{and}\quad
  E\left[ n_\mu \,n_\nu \right] = \M_{\mu\nu}\,,
\end{equation}
where $E[.]$ is the expectation value. We further find
\begin{equation}
  \label{eq:12}
  E[ x_\mu ] = s_\mu\,,\quad\textrm{and}\quad
  E[ x_\mu \,x_\nu ] = \M_{\mu\nu} + s_\mu\,s_\nu\,,
\end{equation}
so the four random variables $x_\mu$ have mean $s_\mu$
and covariances $\M_{\mu\nu}$. 
Using this together with \eqref{eq:8}, we find the expectation value
of the $\F$-statistic as
\begin{equation}
  \label{eq:13}
  E[2\F] = 4 + \rho^2(0)\,,
\end{equation}
in terms of the ``optimal'' signal-to-noise ratio $\rho(0)$, given
by
\begin{equation}
  \label{eq:14}
  \rho^2(0) = s_\mu\,\Minv^{\mu\nu}\,s_\nu = \A^\mu\,\M_{\mu\nu}\,\A^\nu
  = \left(\detVec{s} | \detVec{s}\right)\,.
\end{equation}
Using \eqref{eq:12}, it is straightforward to show that
the quadratic form \eqref{eq:8} can be diagonalized as the sum of 
four squares of Gaussian random variables with unit covariance matrix.
This implies that $2\F$ is distributed according to a (non-central)
$\chi^2$-distribution with \emph{four} degrees of freedom, as
previously shown by \citet{cutler05:_gen_fstat}. The corresponding
non-centrality parameter of this $\chi^2$-distribution is $\rho^2(0)$. 

\subsection{$\F$-statistic of mismatched signals}
\label{sec:f-stat-mism}

In the case of unknown signal parameters, there will generally be an
``offset'' $\Delta\vDoppler$ between the Doppler parameters
$\vDoppler_\sig$ of the signal and the target parameters $\vDoppler$,
i.e.  
\begin{equation}
  \label{eq:19}
  \vDoppler = \vDoppler_\sig + \Delta\vDoppler\,.
\end{equation}
In this case, the projections \eqref{eq:7} can be expressed as
\begin{equation}
  \label{eq:16}
  x_\mu(\vA, \vDoppler_\sig;\vDoppler) = n_\mu(\vDoppler) +
  \A^\alpha \, \R_{\alpha \mu}(\vDoppler_\sig; \vDoppler)\,,
\end{equation}
where the 4x4 matrix $\R_{\alpha\mu}$ is defined as
\begin{equation}
  \label{eq:17}
  \R_{\alpha \mu}(\vDoppler_\sig; \vDoppler) \equiv 
  \left( \detVec{h}_\alpha(\vDoppler_\sig) | \detVec{h}_\mu(\vDoppler) \right)\,.
\end{equation}
This matrix is generally not symmetric, but evidently satisfies the
symmetry relation  
\mb{$\R_{\alpha\mu}(\vDoppler_\sig; \vDoppler) = \R_{\mu\alpha}(\vDoppler;\vDoppler_\sig)$}.
Using this together with \eqref{eq:8}, we can write the
mismatched (SNR)$^2$ as
\begin{eqnarray}
  \rho^2(\vA,\vDoppler_\sig; \vDoppler) &\equiv& E[2\F] - 4 \nonumber\\
  && \hspace*{-3em}= \A^\alpha \R_{\alpha\mu}(\vDoppler_\sig;\vDoppler)\Minv^{\mu\nu}(\vDoppler)
  \R_{\beta\nu}(\vDoppler_\sig; \vDoppler) \A^\beta\,.   \label{eq:18}
\end{eqnarray}
Assuming the target parameters $\vDoppler$ to be ``close'' (in a
suitable sense) to the signal, we can Taylor-expand
these matrices around the signal-location $\vDoppler_\sig$, and
keeping only terms up to second order, we obtain
\begin{eqnarray}
  \label{eq:20}
  \Minv^{\mu\nu}(\vDoppler) &=& \Minv^{\mu\nu}(\vDoppler_\sig) 
  + \partial_i \Minv^{\mu\nu}(\vDoppler_\sig)\,\Delta\Doppler^i\nonumber\\
  & & + {1\over2}\partial_{i j}\Minv^{\mu\nu}(\vDoppler_\sig)\,\Delta\Doppler^i\Delta\Doppler^j\,,\\
  \R_{\mu\nu}(\vDoppler_\sig;\vDoppler) &=& \M_{\mu\nu}(\vDoppler_\sig)
   + \R_{\mu\nu\,i}(\vDoppler_\sig)\,\Delta\Doppler^i\nonumber\\ 
   && + {1\over2} \R_{\mu\nu\,i j}(\vDoppler_\sig)\, \Delta\Doppler^i\,\Delta\Doppler^j\,,
\end{eqnarray}
where indices $i, j$ refer to the Doppler parameters $\Doppler^i$ (with
automatic summation), and where we defined 
\begin{eqnarray}
  \label{eq:21}
  \R_{\mu\nu\,i} &\equiv& \left( \detVec{h}_\mu | \partial_i \detVec{h}_\nu\right)\,,\\
  \R_{\mu\nu\,i j} &\equiv& \left( \detVec{h}_\mu | \partial_{i j}
    \detVec{h}_\nu \right)\,,   \label{eq:21b}
\end{eqnarray}
with \mb{$\partial_i \equiv {\partial / \partial \Doppler^i}$} and
\mb{$\partial_{i j} \equiv {\partial^2 / \partial \Doppler^i
    \partial\Doppler^j}$}.
Substituting these expansions into \eqref{eq:18},
we find to second order
\begin{equation}
  \label{eq:22}
  \rho^2(\Delta\vDoppler) = \A^\mu \left[ 
    \M_{\mu\nu} + \L_{\mu\nu i} \Delta\Doppler^i 
    - \G_{\mu\nu i j} \Delta\Doppler^i  \Delta\Doppler^j
  \right] \A^\nu\,,
\end{equation}
where the first- and second-order coefficients explicitly read as
\begin{eqnarray}
  \L_i &\equiv& \R_i + \trans\R_i + \M\cdot\partial_i\M^{-1}\cdot\M\,,  \label{eq:23}\\
  - \G_{i j} &\equiv& {1\over2}( \R_{i j} + \trans\R_{i j} ) + {1\over2}
  \M\cdot\partial_{i j} \M^{-1}\cdot \M \nonumber\\
  & & + \M\cdot\partial_i \M^{-1}\cdot\trans\R_j + \R_i \cdot\M^{-1}\cdot\trans\R_j\nonumber\\
  & & + \R_i \cdot\partial_j \M^{-1}\cdot\M\,,\label{eq:31}
\end{eqnarray}
using matrix notation, and writing $\trans$ for the transpose of the
\emph{amplitude indices}. 
By applying derivatives to the identity $\M\cdot\M^{-1} = \idm$, we find
\begin{eqnarray}
  \M\!\cdot\!\partial_i \M^{-1}\!\cdot\!\M &=& - \partial_i \M\,, \label{eq:24}\\
  \M\!\cdot\!\partial_{i j}\M^{-1}\!\cdot\!\M &=& -\partial_{i j} \M + 
  2\partial_i \M\!\cdot\!\M^{-1}\!\cdot\!\partial_j \M ,\label{eq:30}
\end{eqnarray}
and using \eqref{eq:25} and the product rule, we further obtain
\begin{eqnarray}
  \partial_i \M &=& \R_i + \trans\R_i \,,  \label{eq:27}\\
  \partial_{i j}\M &=& \R_{i j} + \trans\R_{i j} + 2\, h_{i j}\,, \label{eq:29}
\end{eqnarray}
where we defined
\begin{equation}
  \label{eq:28}
  h_{\mu\nu i j} \equiv \left( \partial_i \detVec{h}_\mu | \partial_j \detVec{h}_\nu \right)\,.
\end{equation}
For simplicity of notation, we assume the Doppler indices in 
\eqref{eq:31} to be implicitly symmetrized, i.e.\ $\G_{ij} = \G_{ji}$,
as any non-symmetric parts will not contribute to the quadratic form \eqref{eq:22}. 
Using the identities \eqref{eq:24}--\eqref{eq:29}, we can reduce the
expansion coefficients \eqref{eq:23}, \eqref{eq:31} to  
\begin{eqnarray}
  \L_i &=& 0\,,   \label{eq:32}\\
  \G_{i j} &=& h_{i j} - \trans\R_i\cdot\M^{-1}\cdot\R_j \,.\label{eq:36}
\end{eqnarray}
The exact vanishing of the first-order coefficient $\L_i$ shows
that the perfectly-matched case, $\vDoppler = \vDoppler_\sig$, is a
local extremum of the $\F$-statistic, as expected.
The zeroth-order term in \eqref{eq:22} corresponds to the
perfectly-matched case \eqref{eq:14}, 
and so we arrive at
\begin{equation}
  \label{eq:33}
  \rho^2(\Delta\Doppler) = \rho^2(0) 
  - \A^\mu\,\G_{\mu\nu\,i j}\,\A^\nu\,\, \Delta\Doppler^i \, \Delta\Doppler^j
  + \O(\Delta\vDoppler^3)\,,
\end{equation}
where $G_{\mu\nu ij}$ can be written more explicitly as
\begin{equation}
  \label{eq:37}
  \G_{\mu\nu i j} = \left( \partial_i \detVec{h}_\mu | \partial_j \detVec{h}_\nu \right)
  - \left( \detVec{h}_\alpha | \partial_i \detVec{h}_\mu \right)\Minv^{\alpha\beta}
  \left( \detVec{h}_\beta | \partial_j \detVec{h}_\nu\right)\,.
\end{equation}
As discussed in \cite{krolak04:_optim_lisa}, this matrix is directly
related to the \emph{projected Fisher matrix} $\overline{\Gamma}$, namely
\begin{equation}
  \label{eq:15}
  \overline{\Gamma}_{ij} \equiv \vA\cdot\G_{ij}\cdot\vA\,,
\end{equation}
describing the statistical information on the Doppler parameters
$\vDoppler^i$, given the amplitude parameters $\A^\mu$. 

\section{The $\F$-statistic metric family}
\label{sec:f-metric}

\subsection{General definitions}
\label{sec:general-definitions}

The relative loss in expected $\F$-statistic due to an offset
$\Delta\vDoppler$ with respect to the signal location $\vDoppler_\sig$
defines a natural dimensionless ``mismatch''\footnote{See
  appendix~\ref{sec:altern-proj-onto} for a discussion of a slightly
  different mismatch definition sometimes found in the literature.} 
$m_\F$, namely 
\begin{equation}
  \label{eq:53}
    m_\F(\vA, \vDoppler_\sig; \Delta\Doppler) \equiv 
    { \rho^2(0) - \rho^2(\Delta\Doppler) \over \rho^2(0)}\,.
\end{equation}
Using \eqref{eq:33} for the mismatched SNR, we can cast the
$\F$-mismatch in the form  
\begin{equation}
  \label{eq:34}
  m_\F = g^\F_{i j}(\vA,\vDoppler) 
  \, \Delta\Doppler^i \Delta\Doppler^j + \O(\Delta\vDoppler^3)\,,
\end{equation}
where we defined the $\F$-statistic metric $g^\F_{ij}$ as
\begin{equation}
  \label{eq:130}
  g^\F_{ij}(\vA,\vDoppler) \equiv 
  { \vA\cdot\G_{i j}(\vDoppler)\cdot\vA \over
    \vA\cdot\M(\vDoppler)\cdot\vA } = \frac{\overline{\Gamma}_{ij}}{\rho^2(0)}\,.
\end{equation}
This expression is identical to that found previously (for the
single-detector case) in \cite{2005LRR.....8....3J,krolak04:_optim_lisa},
referred to as the ``normalized projected Fisher matrix''.
A more elegant method of obtaining this result by projecting the full
parameter-space metric into the Doppler subspace is shown in
appendix~\ref{sec:altern-proj-onto}. 
It is obvious from \eqref{eq:53} that the overall signal amplitude
$h_0$ cancels out, and that the mismatch $m_\F$ is therefore, 
contrary to the Fisher matrix $\overline{\Gamma}_{i j}$, independent of $h_0$.

The Fisher matrix $\overline{\Gamma}_{ij}$ characterizes the
\emph{statistical} uncertainty (due to the presence of noise) of the
maximum-likelihood estimators for the signal parameters. In
particular, the inverse Fisher matrix gives lower bounds on the
variances of the parameter estimators (the so-called Cram\'er-Rao
bound), describing the typical fluctuations by which the location of
the detection-statistic peak will vary in \emph{repeated} experiments. 
The Fisher matrix can therefore be regarded as a measure of the best
possible \emph{accuracy} of parameter estimation.  
The concept of the metric $g^\F_{ij}$, on the other hand, describes
the relative (i.e.\ SNR-independent) ``extent'' of detection-statistic
peaks for any \emph{single} realization of the noise. This gives a
measure of the intrinsic parameter-space \emph{resolution}, which is
not related to the true location of any putative signal. 
However, while conceptually rather different, Eq.~\eqref{eq:130} shows
that there exists a deep connection between the two concepts.

The $\F$-metric \eqref{eq:130} is not a unique metric on the
Doppler-parameter space $\vDoppler$ of the $\F$-statistic, as it
depends on the (generally unknown) signal amplitudes $\A^\mu$. 
Expression \eqref{eq:130} therefore describes a whole family of metrics,
corresponding to different $\vA=\const.$ subspaces of the full
parameter space, which is of very limited direct use for the covering
problem of the Doppler-parameter space. 

However, we can explicitly compute the possible \emph{range} of
mismatches for any given $\vDoppler$ and $\Delta\vDoppler$. 
For this, consider the extrema of $m_\F$ as a function of $\vA$, i.e.  
\begin{equation}
  \label{eq:35}
  0 = {\partial m_\F \over \partial \vA } = 
  {2 \G\vA \over \vA\M\vA } 
  - { \vA\G\vA \over  (\vA\M\vA )^2 } \,2\M\vA\,,
\end{equation}
where we wrote $\G \equiv \G_{i j}\Delta\Doppler^i\Delta\Doppler^j$,
which is a 4x4 matrix in amplitude space. 
Equation \eqref{eq:35} is equivalent to
\begin{equation}
  \label{eq:38}
  \left( \M^{-1}\cdot\G \right)\,\vA = \mhat_\F(\vDoppler,\Delta\vDoppler)\, \vA\,,
\end{equation}
which determines the extremal values $\mhat_\F$ of the $\F$-mismatch
as the \emph{eigenvalues} of $\M^{-1}\!\cdot\!\G$. According to the Rayleigh
principle, the minimum and maximum of the mismatch will be given
respectively by the smallest and largest eigenvalues. As suggested in
\cite{2005LRR.....8....3J}, a more practical
mismatch measure can be constructed from the \emph{mean} of the
eigenvalues, and we can define an ``average'' $\F$-metric as
\begin{equation}
  \label{eq:50a}
  \gbar^\F_{i j}(\vDoppler) \equiv {1\over 4}\, \Tr \left[ \M^{-1}\cdot\G_{i j} \right]\,,
\end{equation}
where the trace $\Tr$ refers to the amplitude indices.
This average $\F$-metric, contrary to $g^\F_{ij}$, is
\emph{independent} of the amplitudes $\A^\mu$, and is therefore of
more practical interest as a metric on the Doppler-parameter
space. The corresponding average $\F$-mismatch $\mbar_\F$ is
simply
\begin{equation}
  \label{eq:137}
  \mbar_\F(\vDoppler,\Delta\vDoppler) \equiv \gbar^\F_{ij}(\vDoppler)\,
  \Delta\vDoppler^i \Delta\vDoppler^j\,.
\end{equation}
We note that in the analogous case of binary-inspiral signals, where
one can equally maximize the detection statistic over some
of the signal parameters (referred to as ``extrinsic'' parameters), the
metric of the reduced parameter space depends again on both extrinsic and
intrinsic parameters \cite{pan04:_physical_templates,buonanno05:_detec}. 
A common choice of mismatch metric for the template placement in
this binary-inspiral context is to use the most conservative case
(based on the concept of the ``minimal match''
\cite{owen96:_search_templates}), namely the worst-case mismatch  
\begin{equation}
  \label{eq:48}
  \mhat_\F^\max(\vDoppler,\Delta\vDoppler) = \max_{_\A}\,
  m_\F(\vA,\vDoppler; \Delta\vDoppler)\,.
\end{equation}
The is referred to as the ``minimax'' prescription in
\cite{pan04:_physical_templates}. 
Contrary to the average metric \eqref{eq:137}, this extremal
metric cannot be expressed as a quadratic form in the
Doppler separations $\Delta\vDoppler$, and so the corresponding
iso-mismatch surfaces are not described by hyper-ellipsoids.

As we will see in the following section, for the type of narrow-band
continuous-wave signals considered here, the 4x4 matrix $\M^{-1}\cdot\G$
has only \emph{two} independent eigenvalues, corresponding directly to
the maximum and minimum possible $\F$-mismatches for any given $\vDoppler$ and
$\Delta\vDoppler$.

\subsection{Narrow-band signals, uncorrelated noise}
\label{sec:neutr-star-spec}

In the following we restrict our analysis to continuous gravitational
waves with a well-defined, slowly varying (intrinsic) frequency $f$. 
This assumption applies, for example, for GWs emitted from
spinning non-axisymmetric neutron stars, and from stellar-mass binary
systems (relevant for LISA). We assume the observation time $T$ to be
much longer than the GW period $1/f_\sig$, such that the number of
cycles $N$ is large, i.e. 
\begin{equation}
  \label{eq:106}
  N = f_\sig\, T \gg 1\,.
\end{equation}
The phase $\phi(t)$ of the signal will be dominated by the zeroth-order
term $2\pi f_\sig t$, while the intrinsic frequency variability and the
Doppler modulations are much smaller corrections.   
Assuming two such narrow-band signals $x(t)$ and $y(t)$, we can
approximate the scalar product \eqref{eq:95} as
\begin{equation}
  \label{eq:41}
  (\detVec{x}|\detVec{y}) \approx T \, S_{\X\Y}^{-1}(f_\sig) \, \av{x^\X\,y^\Y}\,,
\end{equation}
in terms of the time-averaging operator $\av{.}$, defined as
\begin{equation}
  \label{eq:39}
  \av{g} \equiv {1\over T}\int_0^T g(t)\, d t\,.
\end{equation}
Note that the noise matrix $S^{\X\Y}(f_\sig)$ can be considered as a
metric in ``detector space'', allowing us to lower and raise detector
indices, e.g. we could write 
$x_\X(t) \equiv S^{-1}_{\X\Y}(f_\sig) x^\Y(t)$,
and the scalar product \eqref{eq:41} would then read as
$(\detVec{x} | \detVec{y}) = T \,\av{x^\X \,y_\X }$.

For simplicity, we restrict our analysis to the (more
common) case of uncorrelated detector noises, i.e. we assume 
$S^{-1}_{\X\Y} = S^{-1}_\X\,\delta_{\X\Y}$, which allows us to
introduce ``noise weights'' $w_\X$ as 
\begin{equation}
  \label{eq:109}
  w_\X \equiv {S_\X^{-1} \over \Sinv}\,,\quad\textrm{with}\quad
  \Sinv \equiv \sum_\X S_\X^{-1}\,,
\end{equation}
such that $\sum w_\X = 1$. Using this, the scalar product
\eqref{eq:41} can now be expressed as 
\begin{equation}
  \label{eq:110}
  (\detVec{x}|\detVec{y}) =  T\Sinv \,\avS{x\,y}\,,
\end{equation}
where we introduced the \emph{noise-weighted averaging} operator
$\avS{.}$, defined as  
\begin{equation}
  \label{eq:69}
  \avS{Q} \equiv \sum_\X w_\X\,\av{Q^\X}\,.
\end{equation}
The scalar products involved in the expression for the $\F$-statistic
consist of products of slowly varying antenna-pattern functions
\mb{$g(t)\in \{a(t),\,b(t)\}$}, and highly oscillatory, quasi-periodic
functions \mb{$p(t)\in\{\sin^2\phi(t),\,\cos^2\phi(t),\,\sin\phi(t)\cos\phi(t)\}$}.  
In the following we approximate the oscillatory functions as exactly
periodic with period $\tau= 1/f$, i.e.\ $\phi\approx 2\pi\,f\,t$, and
so we can write the time-average $\av{g\,p}$ as 
\begin{equation}
  \label{eq:49}
  \av{g\,p} = \frac{1}{T}\sum_{n=0}^{N-1}
  \int_{t_n}^{t_{n+1}} g(t)\,p(t)\,d t\,,
\end{equation}
where $t_n \equiv n\,\tau$, and $N= T/\tau$ is the number of oscillation cycles.
We can Taylor-expand $g(t)$ in each of the periods as
$g(t) = g(t_n) + \dot{g}(t_n)\,(t - t_n) + \frac{1}{2}\ddot{g}(t_n)(t-t_n)^2+...$,
and using the fact that 
\begin{equation}
  \label{eq:90}
  \frac{1}{\tau}\int_{t_n}^{t_{n+1}} (t-t_n)^m\,p(t)\,dt = \av{t^m\,p}_{\tau}\,,
\end{equation}
where $\av{.}_\tau$ denotes the time-average over one cycle of $p(t)$, 
we obtain the expansion 
\begin{eqnarray}
  \label{eq:92}
  \av{g\,p} &\approx& \frac{\av{p}_{\!\tau}}{T}\, \sum g(t_n)\, \tau +
  \frac{\av{t\,p}_{\!\tau}}{T}\, \sum \dot{g}(t_n)\,\tau + ... \nonumber\\
  &\approx& \av{p}_{\!\tau} \av{g} + \frac{\av{t\,p}_{\!\tau}}{T}
  \left[g(T) - g(0)\right] + ...
\end{eqnarray}
For the averages over one cycle $\tau$ we find:  
\begin{equation}
  \label{eq:123}
  \begin{array}{r c l}
    \av{\sin^2\phi}_\tau  = \av{\cos^2\phi}_\tau  = \frac{1}{2}\,,\quad
    \av{\sin\phi\,\cos\phi}_\tau  = 0\,,\\[0.2cm]
    \av{t^m\sin^2\phi}_\tau \sim 
    \av{t^m\cos^2\phi}_\tau \sim 
    \av{t^m\sin\phi\cos\phi}_\tau \sim \tau^m\,.
  \end{array}
\end{equation}
Under the above assumptions, we can therefore write 
\begin{equation}
  \label{eq:40}
  \av{g\, p} \approx \av{g}\,\av{p} + \O\left({1}/{N}\right)\,,
\end{equation}
and so we can neglect higher-order contributions and
keep only the zeroth-order term $\av{p}\av{g}$ in expressions of this type.
Using this approximation, the 4x4 matrix $\M_{\mu\nu}$, defined in
\eqref{eq:25}, is explicitly found as
\begin{equation}
  \label{eq:54}
  \M_{\mu \nu}  
  \approx {1\over2}\, \Sinv T\, \left( \begin{array}{c c}
      \C & 0 \\
      0 & \C \\
    \end{array}\right)\,,
\end{equation}
where $\C$ is the 2x2 matrix
\begin{equation}
  \C \equiv \left( \begin{array}{c c}
      A & C \\
      C & B \\
    \end{array} \right)\,,
\end{equation}
in terms of the three independent antenna-pattern coefficients 
\begin{eqnarray}
  \label{eq:55}
  A  \equiv \avS{a^2},\quad B \equiv \avS{b^2},\quad C  \equiv \avS{a\,b}\,,
\end{eqnarray}
and we further define $D \equiv A\,B - C^2$.  
Inserting the explicit expressions \eqref{eq:60} for the amplitudes
$\A^\mu$ and using \eqref{eq:54}, we can write the optimal SNR
\eqref{eq:14} explicitly as  
\begin{eqnarray}
  \rho^2(0) &=& \frac{1}{2}\,{h_0^2\, T \,\Sinv}\left[
    \alpha_1\, A + \alpha_2 \, B + 2 \alpha_3\, C \right]\,,\label{eq:42}
\end{eqnarray}
in terms of the amplitudes (writing $\eta \equiv \cosi$):
\begin{equation}
  \label{eq:61}
  \begin{array}{r c l}
    \alpha_1(\eta,\psi) &\equiv& {1\over4} (1+\eta^2)^2 \cos^2 2 \psi + \eta^2 \sin^2 2\psi\,,\\[\vsep]
    \alpha_2(\eta,\psi) &\equiv& {1\over4} (1+\eta^2)^2 \sin^2 2 \psi + \eta^2 \cos^2 2\psi\,,\\[\vsep]
    \alpha_3(\eta,\psi) &\equiv& {1\over4} (1-\eta^2)^2 \sin2\psi \, \cos 2\psi\,.\\
  \end{array}
\end{equation}
We see that the optimal SNR does not depend on the initial
phase $\phi_0$, and it scales linearly with the overall amplitude
$h_0$, and with the square-root of the observation time $T$. The
dependence on the number of detectors will be discussed in the next
section. 

\subsection{Dependence on the number of detectors $\Ndet$}
\label{sec:depend-numb-detect}

A question of central importance is how the SNR and the metric resolution
depend on the number $\Ndet$ of coherently combined detectors.
The dependence of the optimal SNR is very easy to see: in the explicit
expression \eqref{eq:42}, the antenna-pattern coefficients $A,B,C$ are
the noise-weighted averages \eqref{eq:55} over detector-specific
quantities, and the only \emph{scaling} with the number of detectors
$\Ndet$ therefore comes from the total inverse noise floor
$\Sinv = \sum_\X S_\X^{-1}$.  
If we assume, for simplicity, that all $\N$ detectors have a similar
noise floor $S_0$, i.e. $\Sinv \approx \N\,S_0^{-1}$, then the optimal
SNR scales as
\begin{equation}
  \label{eq:50}
  \rho(0) \propto {h_0 \over \sqrt{S_0}}\, \sqrt{T\, \Ndet}\,.
\end{equation}
Doubling the number of detectors (of similar noise floor) therefore
has the same effect on the SNR as doubling the observation time $T$. 

It is not difficult to see from \eqref{eq:110} and definitions
\eqref{eq:25} and \eqref{eq:37}, that both the numerator and
denominator in \eqref{eq:130} have the same scaling with
$\Sinv$ (and therefore $\N$), which cancels out.
To show this more clearly, we write the explicit expression found from
Eq.~\eqref{eq:130} for the multi-detector $\F$-metric, namely
\begin{equation}
  \label{eq:9}
g^\F_{ij} =  \frac{\A^\mu\!\left[ \avS{\partial_i h_\mu\partial_j h_\nu}
    \!-\!\avS{h_\alpha\partial_i h_\nu}\avS{h_\alpha h_\beta}^{\!\!\!-1}
      \avS{h_\beta\partial_j h_\nu} \right]\!\A^\nu}
  {\A^\rho \,\avS{h_\rho h_\sigma} \,\A^\sigma}\,.\nonumber
\end{equation}
It is evident from this expression that the $\F$-metric only depends
on noise-weighted averages of single-detector contributions, but does
not scale with $\Ndet$.   
Note, however, that the multi-detector metric is \emph{not} a simple
average of single-detector metrics.  

Increasing the number of detectors therefore does not increase the
metric resolution in parameter space. This is in strong
contrast to the effect of increasing the observation time $T$, in
which case the metric resolution, and therefore the number of
templates, grows at a high power of $T$ (e.g. $\propto \O(T^5)$, see
\eqref{eq:113}).
We can therefore gain SNR $\propto\sqrt{\Ndet}$ with $\N$
similar-sensitivity detectors, at the cost of ``only'' a linear 
increase $\propto \Ndet$ in the required computing power.
Using a coherent multi-detector search is therefore the
computationally cheapest way to increase SNR in a coherent
wide-parameter search for continuous gravitational waves.

As discussed in Sect.~\ref{sec:general-definitions}, the
metric \emph{resolution} must not be confused with the \emph{accuracy}
of parameter-estimation. The latter increases with SNR (and therefore
also with the number of detectors), as described by the Fisher
information matrix, while the former does not.  
It might still seem somewhat surprising that the additional
``information'' coming from the time delays between detectors does not
result in a higher metric sky resolution.
This can be understood in terms of the diffraction limit, which can be
used to estimate the order of magnitude of the sky resolution
from the ratio of the wavelength $c/f$ to the ``baseline'' $V\,T$, where
$V$ is the orbital velocity. 
The expected metric sky resolution is therefore of the
order $\Delta\Omega_0 \sim c /(T f V)$, which is exactly the metric
scaling found in Sect.~\ref{sec:natur-units-doppl}.   
It is therefore evident that for detector distances of the order
$\sim1,000$~km, the ``integration baseline'' $V\, T$ dominates
the sky resolution starting from observation times as short as
$T\gtrsim 100$~s, and no additional sky resolution can be gained from
the baseline spanned by different detectors. 

\subsection{The $\F$-metric family for high-frequency signals}
\label{sec:f-metric-ground}

In order to explicitly calculate the $\F$-metric family \eqref{eq:9},
we need to consider the derivatives $\partial_i h^\X_\mu$ of the basis 
functions \eqref{eq:135}, namely (omitting detector indices)  
\begin{equation}
  \label{eq:56}
  \begin{array}{r c l}
    \partial_i h_1 &=& \partial_i a\, \cos\phi - a\,\partial_i\phi\,\sin\phi\,,\\[\vsep]
    \partial_i h_2 &=& \partial_i b\, \cos\phi - b\,\partial_i\phi\,\sin\phi\,,\\[\vsep]
    \partial_i h_3 &=& \partial_i a\, \sin\phi + a\,\partial_i\phi\,\cos\phi\,,\\[\vsep]
    \partial_i h_4 &=& \partial_i b\, \sin\phi + b\,\partial_i\phi\,\cos\phi\,.\\
  \end{array}
\end{equation}
The antenna-pattern functions $a(t), b(t)$ do not depend on any of the
Doppler parameters $\vDoppler$ except for the sky position $\vn$. From
their explicit expressions  (cf. Eqs.(12,13) of \cite{jks98:_data})
for \emph{ground-based} interferometers one sees that  
\begin{equation}
  \label{eq:72}
  \partial_\vn a \sim a\sim\O(1),\quad \partial_\vn b \sim b \sim\O(1)\,,
\end{equation}
while the corresponding phase derivatives will typically be of order  
\begin{equation}
  \label{eq:73}
  |\partial_\vn \phi(t)| \sim |\partial_t \phi \, {\vr(t)\over c}|
  \sim 2\pi f\, {r_\orb \over c} + 2\pi f T V/c +...\,.
\end{equation}
Ignoring the constant term, which will not contribute to the
metric, the second term will be much larger than unity if 
the number of cycles $N=f T$ satisfies $N \gg c/V \sim 10^4$. 
This will always be be true for high-frequency signals relevant for
ground-based detectors, for which we can therefore neglect the
antenna-pattern derivatives in \eqref{eq:56} with respect to the phase
derivatives, i.e. 
\begin{equation}
  \label{eq:74}
  \partial_i \phi \gg \partial_i a \,,\; \partial_i b\,,
\end{equation}
This assumption might not hold in the case of low-frequency signals 
that would be more relevant for LISA, or for very short observation 
times. The corresponding calculations are somewhat more tedious, but
lead to equivalent results and are presented in
appendix~\ref{sec:keep-ampl-funct}.   
Using \eqref{eq:74}, \eqref{eq:110} and keeping only the leading-order
terms in \eqref{eq:40}, we can approximate \eqref{eq:28} as 
\begin{eqnarray}
  h_{\mu\nu i j} 
  &\approx& {1\over2 }\Sinv T \,
  \left( \begin{array}{c c c c}
      P^1_{i j} &  P^3_{i j} &  0   &  0   \\
      P^3_{i j} &  P^2_{i j} &  0   &  0   \\
      0         &   0        &  P^1_{i j} &  P^3_{i j} \\
      0         &   0        &  P^3_{i j} &  P^2_{i j} \\
    \end{array}\right)\,,  \label{eq:57}
\end{eqnarray}
with the three independent components
\begin{equation}
  \label{eq:58}
  \begin{array}{c}
    P^1_{i j} = \avS{a^2\,\partial_i\phi\,\partial_j\phi}\,,\quad
    P^2_{i j} = \avS{b^2 \,\partial_i\phi\, \partial_j\phi}\,,\\[\vsep]
    P^3_{i j} = \avS{a\,b\,\partial_i\phi\,\partial_j\phi}\,. \\
  \end{array}
\end{equation}
In the same way, \eqref{eq:21} can be approximated as
\begin{eqnarray}
  \R_{\mu\nu i} 
  &\approx& {1\over2}\Sinv T \, \left( \begin{array}{c c c c}
      0     &  0     &  R^{13}_i & R^{14}_i \\
      0     &  0     &  R^{14}_i & R^{24}_i \\
  -R^{13}_i & - R^{14}_i &    0  &    0     \\
  -R^{14}_i & - R^{24}_i &    0  &    0     \\
  \end{array} \right)\,,  \label{eq:59a}
\end{eqnarray}
where
\begin{equation}
  \label{eq:93}
  \begin{array}{c}
    R^{13}_i = \avS{a^2\,\partial_i \phi} \,,\quad
    R^{24}_i = \avS{b^2\,\partial_i \phi} \,,\\[\vsep]
    R^{14}_i = \avS{a\,b\,\partial_i \phi} \,.
  \end{array}
\end{equation}
Using this and \eqref{eq:54}, we obtain
\begin{equation}
  \label{eq:63}
  \left\{\trans\R_i\M^{-1}\R_j\right\}_{\mu\nu} 
  \approx {1\over 2}\Sinv T
  \left( \begin{array}{c c c c}
      Q_{i j}^1 & Q_{i j}^3  &   0   &  0  \\
      Q_{i j}^3 & Q_{i j}^2  &   0   &  0  \\
      0   &   0 & Q_{i j}^1  & Q_{i j}^3   \\
      0   &   0 & Q_{i j}^3  & Q_{i j}^2   \\
    \end{array}\right)\,,
\end{equation}
with the three independent components
\begin{eqnarray}
  \label{eq:76}
  D\,Q^1_{i j} &=& A \avS{a b \di \phi} \avS{a b \dj \phi}
  + B\avS{a^2\di \phi}\avS{a^2\dj \phi}\nonumber\\
  & & - 2C\avS{a^2\di \phi}\avS{ab\dj \phi}\,,\\[\vsep]
  D\,Q^2_{i j} &=& A\avS{b^2\di\phi}\avS{b^2\dj\phi}
  + B\avS{a b \di \phi}\avS{a b \dj \phi}\nonumber\\
  & & - 2C\avS{ab\di \phi}\avS{b^2\dj\phi}\,,\\[\vsep]
  D\,Q^3_{i j} &=& A\avS{ab\di\phi}\avS{b^2\dj\phi}
  + B\avS{ab\di \phi}\avS{a^2\dj \phi}\nonumber\\
  && \hspace*{-0.7cm}- C\left[\avS{b^2\di\phi}\avS{a^2\dj\phi} 
    + \avS{ab\di\phi}\avS{ab\dj\phi}\right]\,,
\end{eqnarray}
assuming implicit index symmetrization in $i,j$.
Combining this with \eqref{eq:57} and \eqref{eq:37}, 
we find the 4x4 amplitude matrix  
$\G_{\mu\nu}$ in the form
\begin{eqnarray}
  \label{eq:65}
  \G_{\mu\nu} \approx {1\over 2}\Sinv T \,\left( \begin{array}{c c c c}
      m^1 & m^3  &   0   &  0 \\
      m^3 & m^2  &   0   &  0 \\
        0 &  0   &  m^1  & m^3 \\
        0 &  0   &  m^3  & m^2 \\
    \end{array}\right)\,,
\end{eqnarray}
in terms of the three independent mismatch components
\begin{equation}
  \label{eq:66}
  m^r(\vDoppler;\Delta\vDoppler) = m^r_{ij}\,\Delta\vDoppler^i\Delta\vDoppler^j\,,
\end{equation}
where $r \in\{1,2,3\}$, and the corresponding matrices are
\begin{equation}
  \label{eq:136}
  m^r_{ij}(\vDoppler) \equiv P^r_{i j} - Q^r_{i j} \,.
\end{equation}
In analogy to \eqref{eq:42}, we can write the projected Fisher
matrix \eqref{eq:15} as 
\begin{equation}
  \label{eq:115}
  \overline{\Gamma}_{ij} = \frac{1}{2}\,{h_0^2\, T\, \Sinv}\left[
    \alpha_1\, m^1_{ij} + \alpha_2 \, m^2_{ij} + 2 \alpha_3\, m^3_{ij} \right]\,,
\end{equation}
in terms of the amplitudes \mb{$\alpha_r = \alpha_r(\cosi,\psi)$}
defined in \eqref{eq:61}.
This allows us to express the $\F$-metric family \eqref{eq:130} in the
more explicit form 
\begin{equation}
  \label{eq:117}
  g^\F_{i j}(\cosi,\psi;\vDoppler) \equiv  
  { \alpha_1 \, m^1_{i j} + \alpha_2 \, m^2_{i j} + 2 \alpha_3 \, m^3_{i j} 
    \over \alpha_1 \, A + \alpha_2 \, B + 2 \alpha_3 \,C } \,.
\end{equation}
As discussed in Sect.~\ref{sec:general-definitions}, the extrema
$\mhat_\F$ of the mismatch family $m_\F(\vA)$ are given by the
eigenvalues of $\M^{-1}\cdot\G$, namely
\begin{eqnarray}
  \label{eq:67}
  0 &=& \det\left[ \M^{-1}\cdot\G - \mhat_\F \,\idm \right]\nonumber\\
  &=& \det\M^{-1}\, \det\left[ \G - \mhat_\F\,\M \right]\,,
\end{eqnarray}
and therefore
\begin{equation}
  0 = (m^1 - \mhat_\F A)(m^2 - \mhat_\F B) - (m^3 - \mhat_\F C)^2 \,.
\end{equation}
We see that there are maximally \emph{two} independent eigenvalues,
namely  
\begin{equation}
  \label{eq:68}
  \mhat_\F^{\max|\min}(\vDoppler;\Delta\vDoppler) = \mbar_\F \pm \sqrt{ \mbar_\F^2 - \mtilde^2 }\,,
\end{equation}
where
\begin{eqnarray}
  \label{eq:94}
  \mbar_\F &\equiv& (2D)^{-1} \left[B\, m^1 + A\, m^2 -2C\, m^3\right]\,,\\
  \mtilde^2  &\equiv& D^{-1} \left[ m^1\, m^2 - (m^3)^2 \right]\,. \label{eq:94b}
\end{eqnarray}
Note that $\mtilde^2 > 0$ is necessary for the positivity of the
$\F$-mismatch $m_\F$, and is equivalently to the matrix $\G_{i j}$
being positive definite, as can be seen from \eqref{eq:65}.
The extremal solutions \eqref{eq:68} determine the maximum and
minimum possible mismatch, respectively, as well as the average
$\F$-mismatch $\mbar_\F$, for given signal location $\vDoppler$ and
offset $\Delta\vDoppler$.  The corresponding average $\F$-metric is
found from \eqref{eq:137} and \eqref{eq:94} as
\begin{equation}
  \label{eq:131}
  \gbar^\F_{ij}(\vDoppler) = (2D)^{-1}\left[ B \,m^1_{ij} + A \,m^2_{ij} -
    2 C \,m^3_{ij} \right]\,.
\end{equation}
Note that, contrary to the average mismatch $\mbar_\F$, the extremal
values $\mhat_\F^{\min|\max}$ cannot be written as quadratic forms in
the Doppler offsets $\Delta\vDoppler$.
The range of possible $\F$-mismatches for given $\vDoppler$ and
$\Delta\vDoppler$ can be characterized by an intrinsic
``uncertainty'' $\Delta m_\F$, which we define as  
\begin{equation}
  \label{eq:91}
  \Delta m_\F(\vDoppler;\Delta\vDoppler) \equiv {1\over2}\left(\mhat_\F^\max - \mhat_\F^\min\right)
   = \sqrt{ \mbar_\F^2 - \mtilde^2 }\,,
\end{equation}
and we further introduce the relative uncertainty
$\uncertF$ as 
\begin{equation}
  \label{eq:112}
  \uncertF(\vDoppler;\Delta\vDoppler) \equiv {\Delta m_\F \over \mbar_\F} = \left( 1 -
    {\mtilde^2\over \mbar_\F^2 } \right)^{1/2}\,,
\end{equation}
which is bounded in $\uncertF \in [0, 1]$. 
The maximal mismatch $\mhat_\F^\max$ can never be larger than
\emph{twice} the average mismatch $\mbar_\F$, and in most cases it
will be smaller. The average $\F$-metric $\gbar^\F_{ij}$ might therefore be
quite a reliable mismatch measure in practice, which will be confirmed
in the Monte-Carlo studies in Sect.~\ref{sec:appl-isol-puls} for
isolated neutron-star signals. 

\subsection{Long-duration limit: the ``orbital metric''}
\label{sec:long-duration-limit}

Consider the limit of very long observation times compared 
to a day $\tau_d$, i.e. $T \gg \tau_d$. 
We can always write the signal phase $\phi^\X(t)$ at the detector $\X$
as  
\begin{equation}
  \label{eq:77}
  \phi^\X(t) = \phi_\orb(t) + \Delta\phi^\X(t)\,,
\end{equation}
where $\phi_\orb(t)$ is the signal phase modulated by the orbital
motion, while $\Delta\phi^\X(t)$ accounts for the additional diurnal
phase modulation due to the spin of the earth, which is of order 
\begin{equation}
  \label{eq:78}
  \Delta\phi^\X(t) \sim 2\pi\,f\,\Delta\tau^\X(t)\,,
\end{equation}
where the time delay $|\Delta\tau^\X(t)| \lesssim 0.02$~s is periodic over a day.
We restrict ourselves to signals with slowly varying frequency over
the timescale of a day (e.g. excluding neutron stars in close binary systems), 
for which we can assume $\phi_\orb(t)\sim 2\pi f t$.
With these assumptions and \eqref{eq:40}, we can approximate the
typical contributions to the $\F$-metric in the following way:
\begin{eqnarray}
  \label{eq:79}
  \avS{a^2\partial_i\phi} &=& \sum_\X w_\X \av{a^\X a^\X \partial_i \phi^\X}\nonumber\\
  &=& \sum w_\X \av{a^\X a^\X \partial_i \phi_\orb} 
  + \sum w_\X \av{a^\X a^\X \partial_i \Delta\phi^\X}\nonumber\\
  &\approx& \av{\partial_i \phi_\orb} \sum w_\X\av{a^\X a^\X} +
  \avS{a^2\partial_i\Delta\phi}\,.
\end{eqnarray}
According to our assumptions, the average $\av{\partial_i\phi_\orb}$
is at least linear in the observation time $T$, while the second term
will be roughly constant on timescales longer than a day. 
In the limit $T\gg \tau_d$ we therefore find 
\begin{equation}
  \label{eq:81}
  \avS{a^2\partial_i\phi}  \stackrel{T \gg \tau_d}{\approx} A\,\av{\partial_i \phi_\orb}\,.
\end{equation}
In the same way, we can approximate in this limit 
\begin{equation}
  \label{eq:82}
  \begin{array}{c}
    P^1_{i j} \approx A \,\phi^\orb_{i j}\,,\quad
    P^2_{i j} \approx B \, \phi^\orb_{i j}\,,\quad
    P^3_{i j} \approx C \, \phi^\orb_{i j}\,,\\[\vsep]
    Q^1_{i j} \approx A  \,\phi^\orb_i\phi^\orb_j\,,\quad
    Q^2_{i j} \approx  B  \,\phi^\orb_i\phi^\orb_j\,,\\[\vsep]
    Q^3_{i j} \approx  C  \,\phi^\orb_i\phi^\orb_j\,,
  \end{array}
\end{equation}
where we defined
\begin{equation}
  \label{eq:83}
  \phi^\orb_i \equiv \av{\partial_i \phi_\orb}\,,
  \quad\textrm{and}\quad
  \phi^\orb_{i j} \equiv \av{\partial_i\phi_\orb\,\partial_j \phi_\orb}\,.
\end{equation}
Introducing the ``orbital metric'' $g^\orb_{ij}$ as
\begin{equation}
  \label{eq:84}
  g^\orb_{i j} \equiv \phi^\orb_{i j} - \phi^\orb_i \phi^\orb_j\,,
\end{equation}
and the corresponding orbital mismatch $m_\orb$,
\begin{equation}
  \label{eq:85}
  m_\orb \equiv g^\orb_{i j}\, \Delta\Doppler^i\, \Delta\Doppler^j\,,
\end{equation}
we find the limiting $T\gg\tau_d$ approximations
\begin{equation}
  \label{eq:86}
  m^1 \approx A\, m_\orb\,,\;
  m^2 \approx B\, m_\orb\,,\;
  m^3 \approx C\, m_\orb\,.
\end{equation}
The $\F$-mismatch range \eqref{eq:68} therefore reduces to a single
eigenvalue, namely   
\begin{equation}
  \label{eq:87}
  \mhat_\F^{\min|\max} \stackrel{T\gg\tau_d}{\approx} m_\orb\,,
\end{equation}
and the multi-detector $\F$-metric family reduces to the orbital
metric \eqref{eq:84}, i.e.
\begin{equation}
  \label{eq:51}
  g^\F_{ij}(\vA, \vDoppler) \stackrel{T\gg\tau_d}{\approx} g^\orb_{ij}(\vDoppler)\,,
\end{equation}
which is independent of the unknown amplitudes and of the number
and position of detectors. 

Note that a different, but related metric approximation is the 
``phase metric'', which neglects the amplitude modulation $a^\X(t)$,
$b^\X(t)$, but retains the detector-specific phase modulation
$\Delta\phi^\X(t)$, namely
\begin{equation}
  \label{eq:122}
  g^{\phi,\X}_{ij}(\vDoppler) \equiv \av{\partial_i\phi^\X\partial_j\phi^\X}
  - \av{\partial_i\phi^\X}\av{\partial_j \phi^\X}\,.
\end{equation}
While this is intrinsically a single-detector metric, one could
formally generalize it by analogy with the $\F$-metric, and simply
replace the time averages $\av{.}$ by noise-weighted multi-detector
averages $\avS{.}$. In this manner we could define an \emph{ad-hoc}
multi-detector phase metric as
\begin{equation}
  \label{eq:126}
  g^\phi_{ij}(\vDoppler) \equiv \avS{\partial_i\phi\,\partial_j\phi}
  - \avS{\partial_i\phi}\,\avS{\partial_j \phi}\,.
\end{equation}

\section{Application to signals from isolated neutron stars}
\label{sec:appl-isol-puls}

In the following we apply the general framework of the previous
sections to a particular class of gravitational-wave signals, namely 
GWs emitted from isolated spinning neutron stars. 
Restricting ourselves to a specific signal model allows us to
explicitly compute the metric and evaluate different approximations
for various cases of interest. It also allows us to compare the metric
predictions to measured mismatches using a realistic search code for
computing the $\F$-statistic.

\subsection{The phase model}
\label{sec:isol-neutr-star}

GWs emitted from isolated spinning neutron stars can be
described by a very simple phase model, namely
\begin{equation}
  \label{eq:80}
  \phi^\X(t) = 2\pi \, \sum_{k=0}^s {\fkdot(\tauref) \over (k+1)!}\,\left[\tau^\X(t)\right]^{k+1}\,,
\end{equation}
where $\fkdot$ is the $k$-th time derivative of the intrinsic signal
frequency $f(\tau)$ in the solar-system barycenter (SSB), and
$\tau^\X(t)$ is the arrival time in the SSB (with respect to the
reference time $\tauref$) of a wavefront reaching the detector
$\X$ at time $t$. 
Neglecting relativistic effects, this relation is simply given by
\begin{equation}
  \label{eq:64}
  \tau^\X(t) = t + {\vr^\X(t)\cdot \vn \over c}  - \tauref \,,
\end{equation}
where $\vr^\X(t)$ is the position of detector $\X$ with respect to
the SSB. This can be separated into an orbital and a spin component,
namely  
\begin{equation}
  \label{eq:70}
  \vr^\X(t) = \vr_\orb(t) + \vr_\spin^\X (t)\,,
\end{equation}
where $\vr_\orb$ is the position of the earth in the SSB, and
$\vr_\spin^\X$ is the position of the detector on earth. 
The unit vector $\vn$ denotes the sky position of the source, which
can be written as
\begin{equation}
  \label{eq:89}
  \vn = \left(\cos\delta\, \cos\alpha , \cos\delta\, \sin\alpha, \sin\delta \right)\,,
\end{equation}
in terms of the equatorial coordinates \emph{right ascension} $\alpha$ and
\emph{declination} $\delta$.
For this phase model we have the derivatives
\begin{eqnarray}
  \label{eq:88}
  {\partial \phi^\X(t) \over \partial \fkdot} &=& 2\pi
  {\left[\tau^\X(t)\right]^{k+1}\over (k+1)!}\,,\\
  {\partial \phi^\X(t) \over \partial \vn} &=& 2\pi {\vr^\X(t)\over c}\,
  \sum_{k=0}^s {\fkdot \over k!}\left[ \tau^\X(t) \right]^k\,.  \label{eq:88b}
\end{eqnarray}
The intrinsic signal frequency $f(\tau)$ can usually be assumed to be
a slowly varying function of time, and typically one only needs to
include a small number of spindown coefficients $\fkdot$, between
$s=0$ and $s=3$, say.    
With these phase derivatives and the explicit antenna-pattern functions
$a(t)$, $b(t)$ (Eqs.(12),(13) in \cite{jks98:_data}), we can
numerically compute the different metrics derived in
Sect.~\ref{sec:f-metric}. 

The corresponding time integrals involved in these expressions were
computed numerically using a Gauss-Legendre quadrature of order
2000. This results in a (saturated) convergence of the metric
components $g_{ij}$ and the antenna-pattern integrals $A,B,C$ at the
level of a relative precision of $10^{-10}$, while the corresponding
mismatches, $m = g_{ij}\Delta\Doppler^i\Delta\Doppler^j$, only
converge to a level of about $10^{-5}$. 
The weaker convergence of $m$ is can be attributed to the fact that
the metrics are highly ill-conditioned matrices, as will be discussed
further in Sect.~\ref{sec:metr-determ-eigenv}.

\subsection{Natural units for the Doppler offsets $\Delta\Doppler$}
\label{sec:natur-units-doppl}

We can easily find the scaling and the order of magnitudes of the metric
components: keeping only the dominant scaling-terms in the phase
derivatives \eqref{eq:88}, \eqref{eq:88b} we find 
\begin{equation}
  \label{eq:114}
  \partial_{f^{(k)}} \phi^\X \sim t^{k+1}\,,\quad
  \partial_\vn \phi^\X \sim f \, \vr_\orb(t) /c\,.
\end{equation}
Taylor-expanding $\vr_\orb(t)$ for ``short'' observation times  $T\ll 1$~year, 
and neglecting the antenna-pattern functions $a(t)\sim b(t) \sim \O(1)$, we
find the following scaling for the metric components \eqref{eq:122}:
\begin{equation}
  \label{eq:116}
  \begin{array}{rcl}
    g_{f^{k}f^{k'}} &\sim& T^{k+1}\,T^{k'+1}\,,\\[\vsep]
    g_{n^i n^j} &\sim& (f V /c)^2\,T^2\,e_i\,e_j\,,\\[\vsep]
    g_{f^{k} n^i} &\sim& T^{k+1}\, f T V/c \,\, e_i\,,
  \end{array}
\end{equation}
where $e^i$ is the unit vector along the orbital velocity $V^i$.
These estimates allow us to read off the (minimal) scaling of the
metric determinant $g=\det g_{ij}$ and of the corresponding volume measure
$\sqrt{g}$, namely
\begin{equation}
  \label{eq:113}
  \sqrt{g} \sim f^2 \,\O(T^2) \, \prod_{k=0}^s T^{k+1}\,,
\end{equation}
where $s$ is the number of spindown coefficients to include.
Note that the dominant term \eqref{eq:116} of the sky metric 
$g_{n^i n^j}$ is degenerate and will only be regularized by higher order
terms. Therefore the scaling of the number of sky templates will be
\emph{at least} of the order $\O(T^2)$, but is likely to be higher.
If we consider the case of a single spindown, i.e.\ $s=1$, we find
the scaling for the volume (and therefore number of templates) of the
order $\sqrt{g}\sim f^2 \,\O(T^5)$.  

In order for these metric components to be dimensionless and of order
unity, we should measure the Doppler offsets $\Delta\vDoppler^i$  
in terms of ``natural units'' $\Delta\vDoppler_0^i$, namely
\begin{equation}
  \label{eq:125}
\Delta f_0^{(k)} \equiv 1 / T^{k+1}\,,\quad
\Delta n_0 \equiv c / (f\,T\,V)\,,
\end{equation}
where $V/c \approx 10^{-4}$. We denote the Doppler offsets
in natural units as  
\begin{equation}
  \label{eq:119}
  \Delta\widehat{\vDoppler}^i \equiv \Delta\vDoppler^i / \Delta\vDoppler_0^i\,.
\end{equation}
Note that $\vn$ is the unit vector pointing to a sky location, and using its
expression \eqref{eq:89} in terms of equatorial coordinates $\alpha$,
$\delta$, we find 
\begin{equation}
  \label{eq:120}
  |\Delta\vn|^2 = \cos^2\delta\,(\Delta\alpha)^2 + (\Delta\delta)^2\,.
\end{equation}
For small offsets, $|\Delta\vn|$ is simply the angular
distance corresponding to offsets $\Delta\alpha$ and
$\Delta\delta$. We therefore define the natural-unit angular
offset $\dOm$ on the sky as 
\begin{equation}
  \label{eq:121}
  \dOm \equiv {|\Delta\vn| \over \Delta n_0 }\,.
\end{equation}
In these units, the statement of ``small'' Doppler offsets
in the expansion \eqref{eq:33} is really meaningful, i.e. the validity
of the metric approximation is guaranteed in the regime
\begin{equation}
  \label{eq:124}
  \Delta \widehat{f}^{(k)} \ll 1\,,\quad
  \dOm \ll 1\,.
\end{equation}
However, as will be seen later, this condition is in many cases not
necessary for the approximate validity of the metric. 
Note that the metric iso-mismatch ellipses on the sky are highly
anisotropic (the dominant contribution shown in \eqref{eq:116} is in fact
degenerate), and the actual metric scale can therefore deviate
significantly from $\dOm$, depending on the direction of the 
angular offset. In fact, $\dOm$ corresponds to the larger of the
two angular eigenvalues on the sky (resulting in the smaller
mismatch-scale), but it nevertheless captures the correct scaling of
the sky metric.

\subsection{Numerical exploration of the parameter space}
\label{sec:monte-carlo-expl}

The results in the following sections are based on Monte-Carlo
simulations of the parameter space of signals and Doppler
offsets. Here we summarize how the underlying
random distributions are generated.  
For the signal amplitudes $\A^\mu$, given in \eqref{eq:60}, we pick
$\cosi$ and $\psi$ with uniform distributions from the ranges 
\mb{$\cosi \in [ -1, 1 ]$} and \mb{$\psi \in [ 0, \pi ]$}.
We simply fix the overall amplitude to $h_0=1$ and the initial phase
to $\phi_0=0$, as these parameters have no effect on the metric.  
The Doppler parameters $\vDoppler^i$ of the signal are picked (with
uniform distributions) from the ranges 
\mb{$f \in [ 100, 200 ]$~Hz}, \mb{$\fdot \in [ -10^{-9}, 10^{-9} ]\;s^{-2}$},
\mb{$\alpha \in [ 0, 2\pi ]$} and \mb{$\sin\delta \in [-1, 1]$}.
Note that $\delta$ is picked in such a way as to obtain an 
\emph{isotropic} distribution of points $\{\alpha, \delta\}$ on the
sky sphere.  

The appropriate selection of random Doppler offsets $\Delta\vDoppler^i$ is
more delicate, because their distribution should be roughly
``isotropic'' in some metric sense, and the corresponding
$\F$-mismatches should lie in a ``reasonable'' range, where the
metric approximation is applicable, e.g. \mb{$m_\F \le 0.5$} say. 
Note that the second requirement is irrelevant for the purpose of
studying intrinsic properties of the metric and for comparing
different metric approximations with each other. However, it is
essential for a comparison of the metric predictions to 
\emph{measured} $\F$-mismatches, which will be described in
Sect.~\ref{sec:comp-meas-snr}.  

We use the following algorithm to generate suitable Doppler
offsets $\Delta\vDoppler^i$, satisfying the above requirements:   
\begin{enumerate}
\item Pick the signal parameters $\A^\mu$ and $\vDoppler^i$ as
  described above, and compute the corresponding $\F$-metric
  $g^\F_{ij}(\cosi,\psi;\vDoppler)$ at this parameter-space point. 

\item Pick a random offset vector $\Delta\widehat{\Doppler}^{'i}$ in
  natural units with a uniform distribution in the hypercube, i.e. 
  $\Delta{\widehat{\Doppler}}^{'i}\in[-1,1]$. The corresponding
  dimensional offsets $\Delta{\vDoppler'}^i$ are therefore uniformly
  distributed in  $\Delta f' \in [-\Delta f_0,\Delta f_0]$, 
  $\{\Delta{\alpha'},\Delta{\delta'}\} \in [-\Delta n_0,\Delta n_0]$, 
  and $\Delta\fdot' \in [-\Delta \fdot_0, \Delta \fdot_0 ]$

\item Normalize $\Delta{\vDoppler'}^i$ using the metric, i.e. construct
 $\dopVec{e}^i_\Delta \equiv \Delta{\Doppler'}^i/|\Delta{\Doppler'}|$,
 where $|\Delta{\Doppler'}|=
 \sqrt{g^\F_{ij}\,\Delta{\Doppler'}^i\Delta{\Doppler'}^j}$ 

\item Pick a distance $d\equiv\sqrt{m}$ with uniform distribution
  in $d\in[0.1,\sqrt{0.5}]$. 

\item The resulting Doppler offset is then obtained as 
  $\Delta\vDoppler^i = d\,\,\dopVec{e}^i_\Delta$.

\end{enumerate}
The lower bound of $m_\F \ge 0.01$ on the generated $\F$-mismatches is
chosen in order to avoid numerical errors dominating the relative
differences between mismatches, which happens especially when
comparing to measured mismatches.
We typically generate about 250,000 random choices of signal
parameters and offsets, for observation times ranging from 
$T=12$~hours to $T=200$~hours in steps of 4 hours, which corresponds
to about 5000 trials for each choice of observation time.
By construction, the distribution of $m_\F$ is such that $\sqrt{m_\F}$
is uniformly distributed in the range $[0.1, \sqrt{0.5}]$. 
The distribution of the Doppler offsets
$\Delta\widehat{\vDoppler}$ in natural units is found to be approximately Gaussian
with zero mean and a standard deviation of about $0.3$. There are,
however, a few percent of offsets reaching up to a few hundreds in
natural units, which corresponds to highly degenerate directions of
the metric.

\subsection{Intrinsic uncertainty of the $\F$-metric}
\label{sec:intr-uncert-f}

As discussed in Sect.~\ref{sec:general-case}, the reduced parameter
space of the $\F$-statistic is the Doppler space $\vDoppler$, while
the amplitude space $\vA$ has been ``projected out'' by maximization.
For the problem of covering the Doppler-parameter space with
templates, the amplitudes represent \emph{unknown} external
parameters, and so we cannot directly use the $\F$-metric
family $g^\F_{ij}(\vA,\vDoppler)$ for the covering problem.  
However, as shown in Sect.~\ref{sec:neutr-star-spec}, we can compute
an average $\F$-metric $\gbar^\F_{ij}(\vDoppler)$, and a
corresponding intrinsic relative uncertainty
$\uncertF(\vDoppler,\Delta\vDoppler)$, which are both independent of
the unknown amplitudes $\vA$.  
\begin{figure}[h!tbp]
  \includegraphics[width=0.5\textwidth,clip]{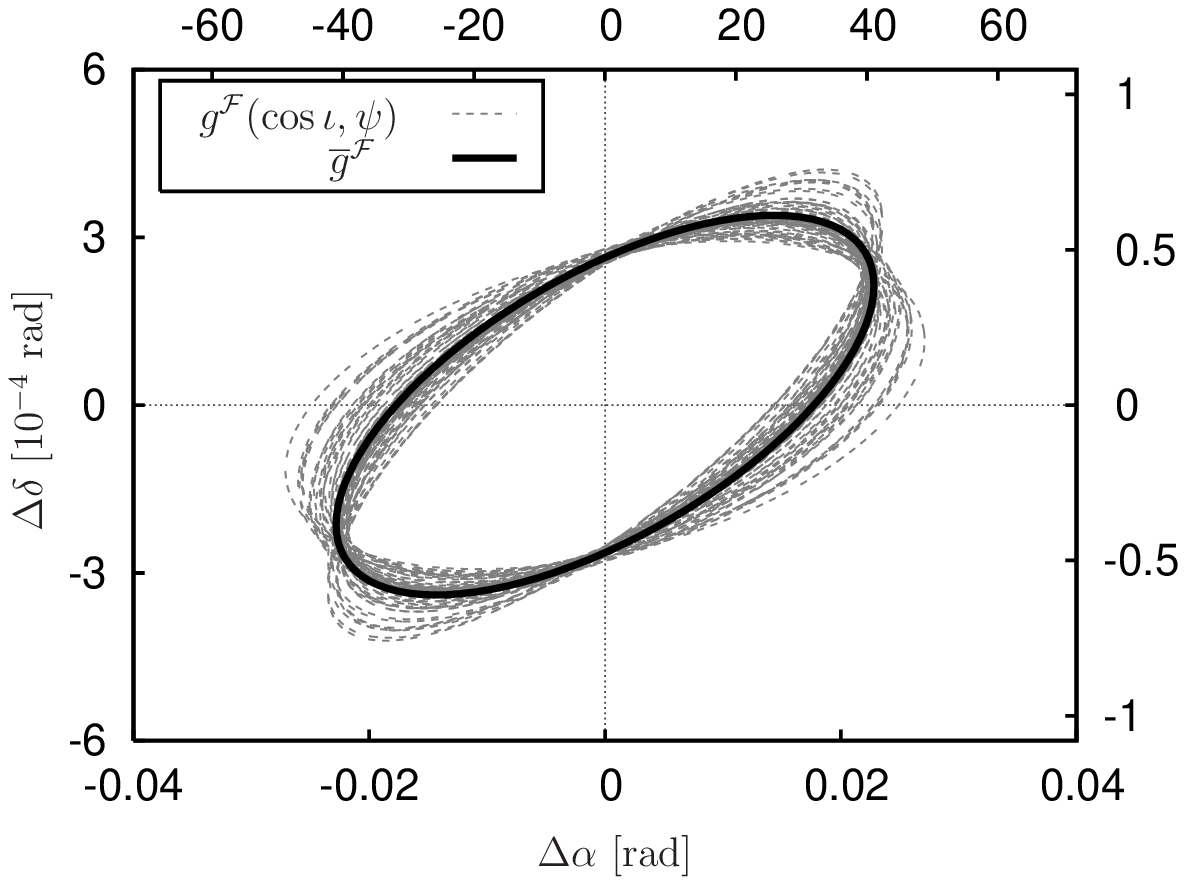}
  \caption{Iso-mismatch ($m=0.1$) ellipses of the $\F$-metric family
    $g^\F_{ij}(\cosi,\psi;\vDoppler)$ for 100 random choices 
    $\{\cosi,\psi\}$, and of the average metric $\gbar^\F_{ij}(\vDoppler)$. 
    Parameter-space cut along the sky plane $\{\alpha,\delta\}$.
    The top and right-hand axes show the corresponding offsets in
    natural units. 
    [Parameters: $f=100$~Hz, $\alpha=1.45~\rad$, $\delta = 0~\rad$, $\fdot=0$,
    detector = 'L1', GPS start-time $t_0 = 810720013$~s, duration
    $T=50$~hours.]  
  } 
  \label{fig:gF_ellipses}
\end{figure}
\begin{figure}[h!tbp]
  \includegraphics[width=0.5\textwidth,clip]{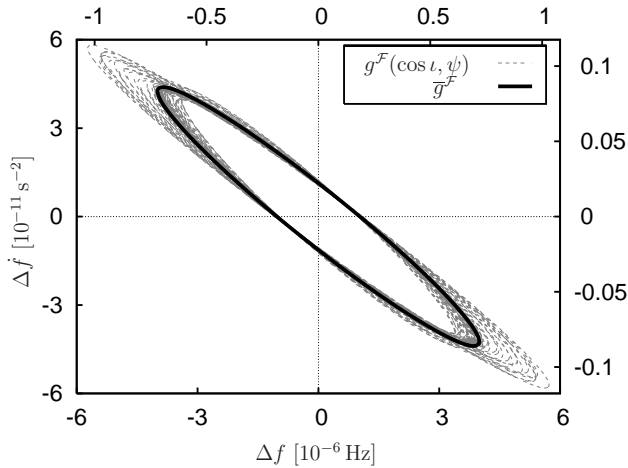}
  \caption{Same as Fig.~\ref{fig:gF_ellipses} for a parameter-space
    cut along the $\{f,\fdot\}$ plane.
    The top and right-hand axes show the corresponding offsets in
    natural units. } 
  \label{fig:gF_ellipses2}
\end{figure}
This uncertainty range in the predicted mismatch is illustrated in
Fig.~\ref{fig:gF_ellipses} and Fig.~\ref{fig:gF_ellipses2}, which show
different parameter-space cuts through the iso-mismatch hyper-ellipsoids
for the average metric $\gbar^\F_{ij}$ and for 100 randomly 
picked members of the $\F$-metric family $g^\F(\cos\iota,\psi)$.
Note that the \emph{intersection} of all these mismatch
hyper-ellipsoids $m_\F<0.1$ corresponds to the worst-case mismatch
region  $\mhat_\F^\max < 0.1$, which is clearly not a hyper-ellipsoid. 
\begin{figure}[h!tbp]
  \includegraphics[width=0.48\textwidth,clip]{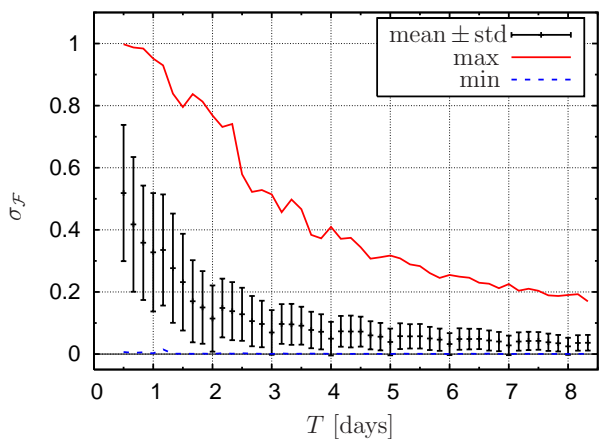}
  \caption{Distribution of the relative intrinsic uncertainty
    $\uncertF$ of the $\F$-mismatch as a function of observation time
    $T$, for the single-detector case. 
    Plotted are the mean, standard deviation and extremal values of the  
    distribution for each value of $T$.
  }
  \label{fig:intr_uncertainty_mF}
\end{figure}
Fig.~\ref{fig:intr_uncertainty_mF} shows the distribution of the
relative uncertainty $\uncertF(\vDoppler,\Delta\vDoppler)$, as a
function of the observation time $T$ for randomly chosen signal
parameters and offsets (using the algorithm described in
Sect.~\ref{sec:monte-carlo-expl}).   
These results are for the single-detector case only, the corresponding
dependence on the number of detectors is investigated in 
Sect.~\ref{sec:f-mismatch-m_f}.  
We see in Fig.~\ref{fig:intr_uncertainty_mF} that for short
observation times, of the order of $T\sim12$~hours, the relative
uncertainty can be as large as $100\%$. For longer observation times
of the order of a few days, $\uncertF$ decreases substantially, 
but even for $T\sim8$~days, the intrinsic uncertainty can still reach
up to about $20$~\%.

\subsection{Comparing different metric approximations}
\label{sec:prop-m_f-isol}

For the following comparisons of different metric approximations, it
will be useful to define the \emph{relative error} $\relError(a,b)$
between two quantities $a$ and $b$ as 
\begin{equation}
  \label{eq:129}
  \relError(a, b) \equiv { a - b \over (|a| + |b|)/2 }\,,
\end{equation}
which is the harmonic mean of $(a-b)/|a|$ and $(a-b)/|b|$. This
definition has the advantage of being bound within
$[-2, 2]$ even for large deviations $a\gg b$ or $b \gg a$, while it
agrees with the more common definitions of relative errors for small
values $|\relError| \ll 1$.

\subsubsection{Monte-Carlo study of mismatches}

As we have seen in Fig.~\ref{fig:intr_uncertainty_mF}, the intrinsic
relative uncertainty $\uncertF$ decreases substantially on the timescale of a
few days, corresponding to a convergence $g^\F_{ij}\rightarrow \gbar^\F_{ij}$.
The average metric $\gbar^\F_{ij}$ is therefore expected to be an
increasingly reliable approximation of the full $\F$-metric
$g^\F_{ij}$ with longer observation times $T$.
\begin{figure}[h!tbp]
  \centering
  \includegraphics[width=0.5\textwidth,clip]{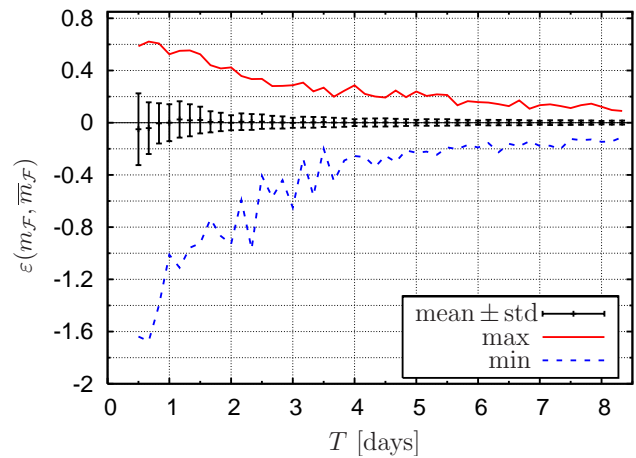}
  \caption{Distribution of relative errors $\relError(m_\F,\mbar_\F)$
    of the (single-detector) average $\F$-mismatch $\mbar_\F$ with
    respect to the exact $\F$-mismatch $m_\F$ as a function of
    observation time $T$.
    Plotted are the mean, standard deviation and extremal values of
    the distribution for each value of $T$.
  }  
  \label{fig:quality_mFav}
\end{figure}
This is indeed the case, as can be seen in Fig.~\ref{fig:quality_mFav},
which shows the distribution of relative errors
$\relError(m_\F,\,\mbar_\F)$ as a function of observation time $T$.
This distribution was obtained using random choices of the signal
parameters and Doppler offsets as described in
Sect.~\ref{sec:monte-carlo-expl}.
The standard deviation of the relative errors is about $~20$~\% for
short observation times $T\lesssim1$~day, but rapidly decreases to
about $1$~\% for timescales of a few days.
Note that these errors are somewhat smaller than could have been
expected from the intrinsic uncertainty $\uncertF$ shown in
Fig.~\ref{fig:intr_uncertainty_mF}, which indicates that mismatches
near the extremal values $m_\F^{\min|\max}$ are less likely than those
closer to the average $\mbar_\F$.  

As shown in Sect.~\ref{sec:long-duration-limit}, the $\F$-metric
family $g^\F_{ij}$ converges to the orbital metric $g^\orb_{ij}$ for
very long observation times $T\gg1$~day, but it is not obvious on
which timescales this convergence happens in practice. 
\begin{figure}[h!tbp]
  \includegraphics[width=0.5\textwidth,clip]{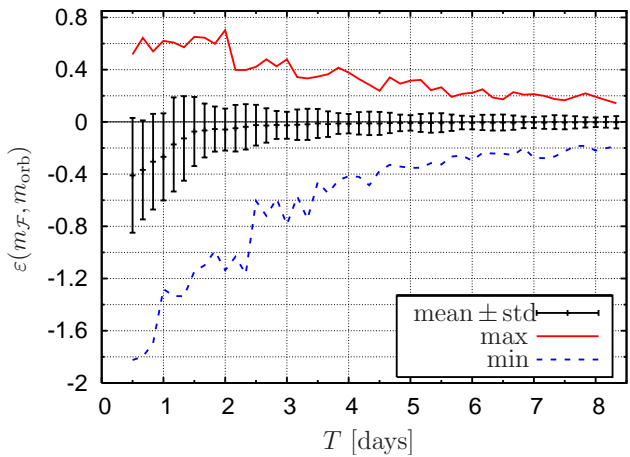}
  \caption{Distribution of relative errors $\relError(m_\F,m_\orb)$
    of the orbital metric  with respect to the (single-detector)
    $\F$-metric as a function of observation time $T$.  
    Plotted are the mean, standard deviation and extremal values of the  
    distribution for each value of $T$.
  }     
  \label{fig:quality_mOrb}
\end{figure}
Fig.~\ref{fig:quality_mOrb} shows the relative errors 
 $\relError(m_\F,m_\orb)$ of the orbital metric compared to the exact
 $\F$-metric, as a function of observation time.  
Note that a plot with the distribution of $\relError(m_\F,m_\phi)$
would look virtually indistinguishable. 
\begin{figure}[h!tbp]
  \includegraphics[width=0.5\textwidth,clip]{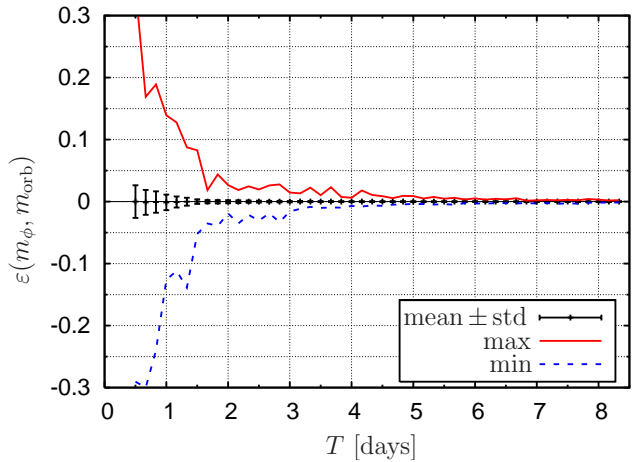}
  \caption{Distribution of relative errors $\relError(m_\phi,m_\orb)$
    of the orbital metric  with respect to the
    phase metric as a function of observation time $T$. 
    Plotted are the mean, standard deviation and extremal values of the  
    distribution for each value of $T$.
  }     
\label{fig:diff_mOrb_mPh}
\end{figure}
These two metric approximations perform nearly identically as far as
predicting mismatches is concerned, as can be seen in
Fig.~\ref{fig:diff_mOrb_mPh} showing the distribution of relative
errors $\relError(m_\phi,m_\orb)$.
Note that the orbital as well as the phase metric show a tendency to
\emph{overestimate} the mismatch for short observation times
$T\lesssim 1$~days, which is apparent in Fig.~\ref{fig:quality_mOrb}.  
Summarizing, we conclude that the ``mismatch quality'' of the orbital
and of the phase metric seem virtually identical, and both
approximations perform only sightly worse than the average $\F$-metric
shown in Fig.~\ref{fig:quality_mFav}.  
However, as will become clear in the next section, there is an important
difference between the phase metric and the orbital metric, which does
not manifest itself in this Monte-Carlo study due to the smallness of
the relevant parameter space in which the difference becomes apparent. 

\subsubsection{Metric determinants and eigenvalues}
\label{sec:metr-determ-eigenv}

\begin{figure}[h!tbp]
  \centering
  \includegraphics[width=0.5\textwidth]{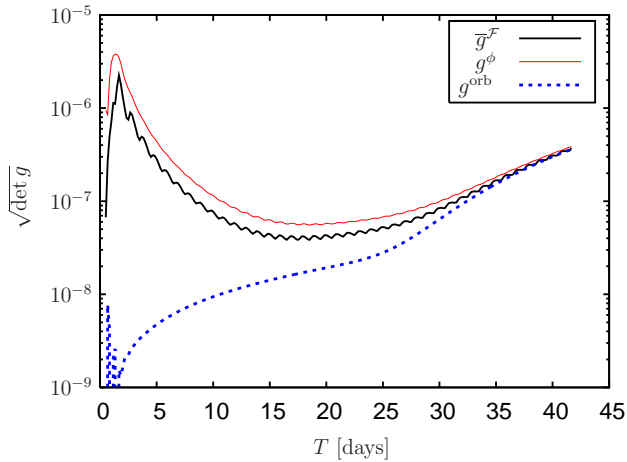}
  \caption{Determinants of different metric approximations (in natural
    units) as functions of observation time $T$.
    [Parameters: $f = 100$~Hz, $\fdot=0$, $\alpha=1.0~\rad$,
    $\delta=0.5~\rad$, detector = 'H1', GPS start-time $t_0 = 792576013$~s]
  }
  \label{fig:metric_determinants}
\end{figure}

\begin{figure*}[htbp]
  \centering
  \mbox{\includegraphics[width=0.45\textwidth,clip]{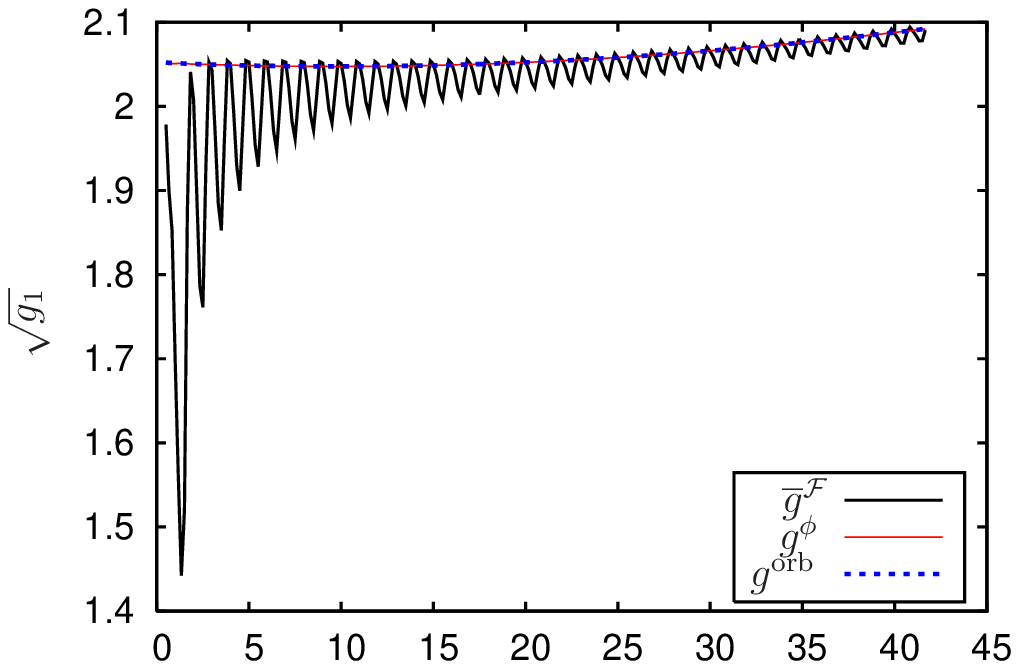}
    \hspace*{-0.5cm}\includegraphics[width=0.45\textwidth,clip]{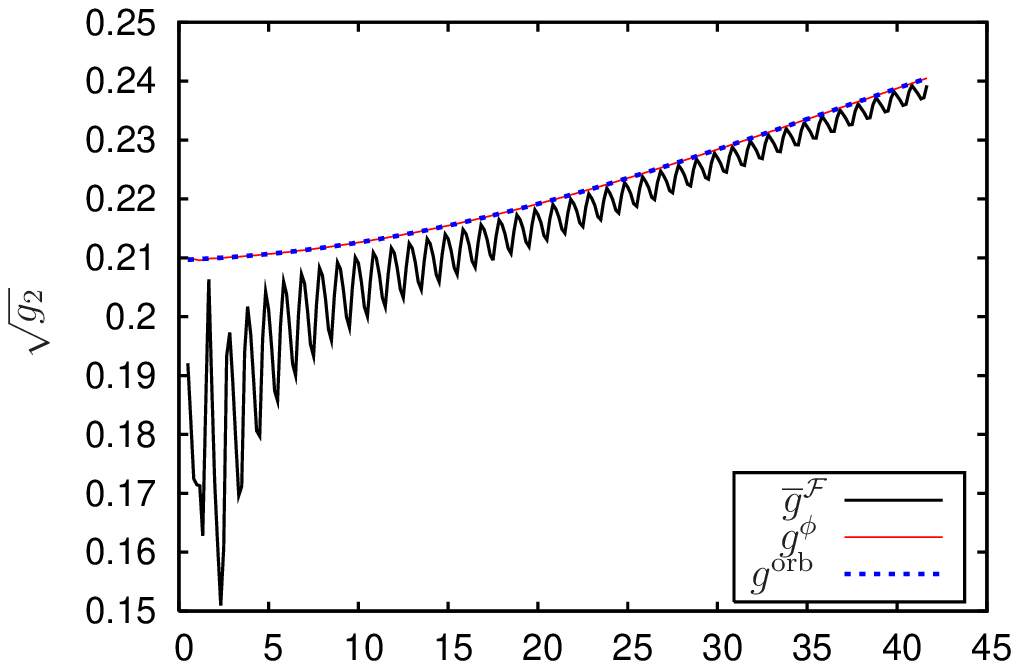}}\\[-0.6cm]
  \mbox{\includegraphics[width=0.45\textwidth,clip]{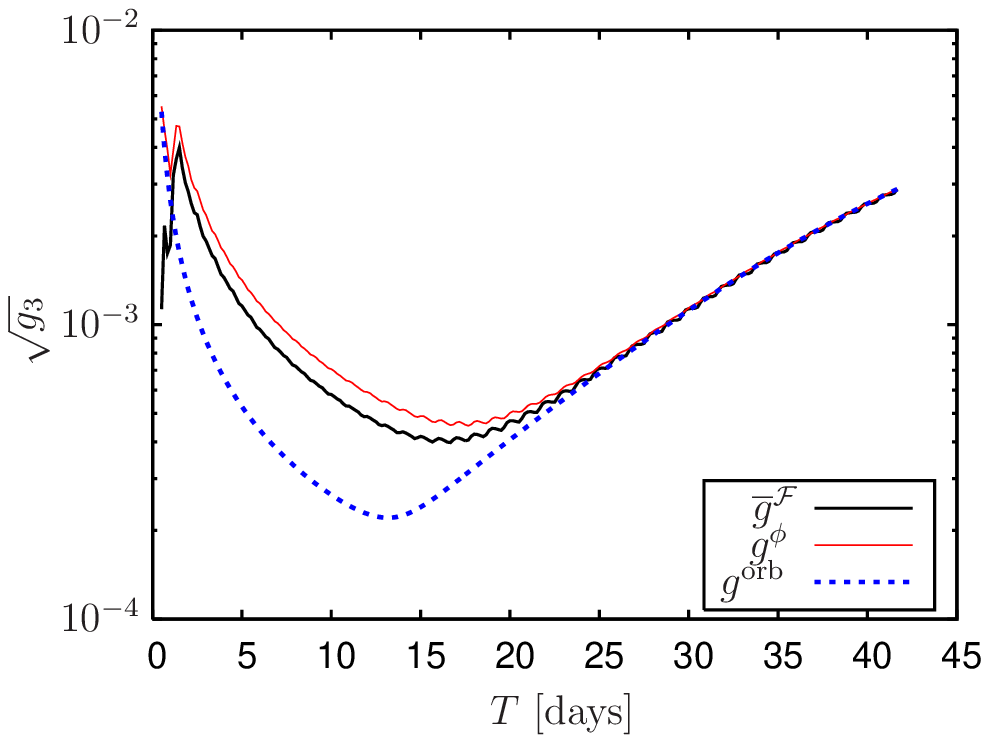}
    \hspace*{-0.5cm}\includegraphics[width=0.45\textwidth,clip]{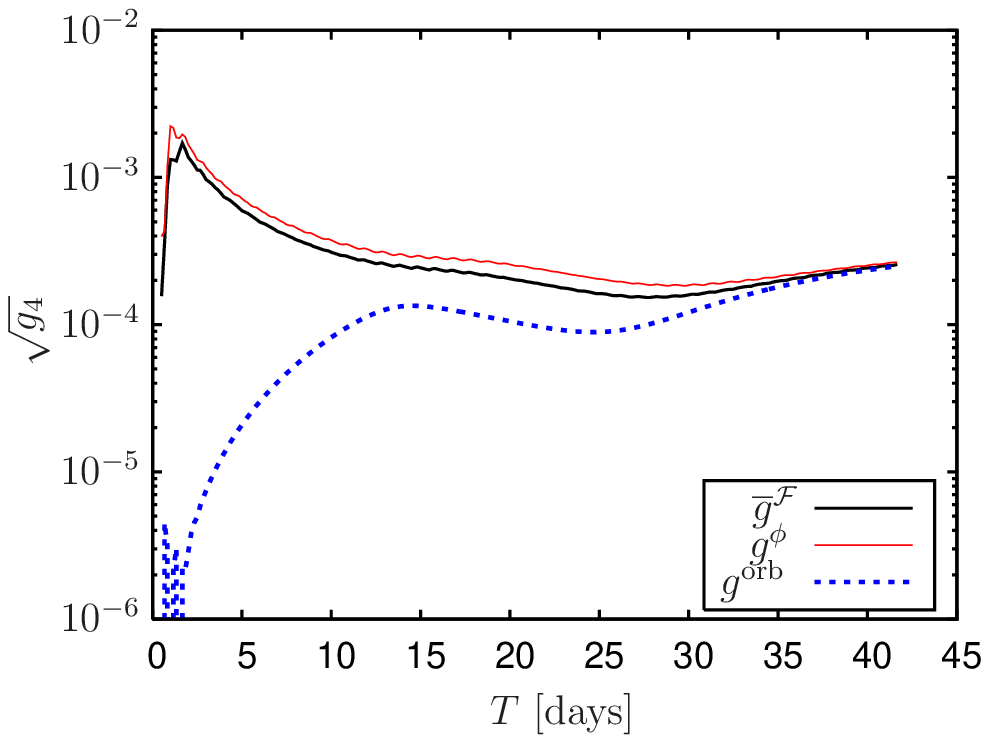}}
  \caption{Square roots of the four eigenvalues $g_i$ of different
    metric approximations (in natural units) as functions of
    observation time $T$. [Same parameters as in
    Fig.~\ref{fig:metric_determinants}]}    
  \label{fig:metric_eigenvalues}
\end{figure*}

A complementary way of comparing different metrics is to look at their
eigenvalues and determinants. These invariant quantities characterize
the semi-major axes and the volume of iso-mismatch hyperellipsoids,
which are important properties for the template covering of the
parameter space.   
We denote the four metric eigenvalues as 
$g_1 \ge g_2 \ge g_3 \ge g_4$.
The metric for isolated neutron-star signals is quite generally
represented by highly ill-conditioned matrices in the default
parameter-space coordinates ($\alpha,\delta,f, \fdot,...$).   
That means that $g_1$ is typically many orders of magnitude larger
than $g_4$, corresponding to very thin and elongated mismatch
hyperellipsoids. This property can be characterized by the so-called
\emph{condition number}, defined as $\kappa \equiv {g_1 / g_4}$.
Well-conditioned matrices have $\kappa \sim\O(1)$, while the metrics
encountered here typically have $\kappa(g^\F_{ij}) \sim 10^{20}$ in
SI units, and $\kappa(g^\F_{\widehat{ij}})\sim 10^8$
in natural units (cf. Sect.~\ref{sec:natur-units-doppl}).
Ill-conditioned matrices can strongly amplify numerical errors in 
computing their maps, inverses, determinants or eigenvalues. 
One has to be very careful when handling such matrices numerically, as
the results can be quite unreliable. 
Using SI units, it turns out to be all but impossible to compute the
determinant or the eigenvalues using standard double-precision
arithmetic, which suffers from complete loss of significant digits in
this case. 
In natural units, however, these quantities can be computed (albeit
not with good precision, and not for all points in parameter space),
and the results for one example are shown in
Fig.~\ref{fig:metric_determinants} and
Fig.~\ref{fig:metric_eigenvalues}.  
Contrary to the apparent fast convergence of the orbital- and the
$\F$-mismatches shown in Fig.~\ref{fig:quality_mOrb}, we see 
a very different picture for the metric determinants in
Fig.~\ref{fig:metric_determinants}.    
For timescales of a few days, the determinant of the orbital metric is
a few orders of magnitude smaller than the determinants of the other
metric approximations, and the expected convergence \eqref{eq:51}
takes place only for timescales longer than $T\gtrsim 1$~month.   

How can this be reconciled with the apparently much faster convergence
of the mismatches in Fig.~\ref{fig:quality_mOrb}?
In order to better understand this, let us look at the four
eigenvalues $g_i$ as functions of $T$, which is shown in
Fig.~\ref{fig:metric_eigenvalues}.   
We see that the largest discrepancy of the orbital metric occurs for
the smallest eigenvalue, $g_4$, corresponding to the most degenerate
direction of the metric. The largest two eigenvalues $g_1, g_2$
agree well, and $g_3$ only differs by a factor of a few.
In order for the mismatch to be affected by the smallest eigenvalue,
we would have to pick a Doppler offset that is \emph{very} 
closely aligned with the most degenerate principal axis, as any
appreciable offset along the other axes would easily dominate the
total mismatch.  
A very rough order-of-magnitude estimate of the probability $p$ of
picking such a direction yields an upper bound of   
$p \lesssim \sqrt{(g_4/g_1)(g_4/g_2)(g_4/g_3)} \lesssim 10^{-6}$.
It is therefore very unlikely to pick a Doppler offset 
for which the mismatch is dominated by the smallest eigenvalue.
This is consistent with the fact that in about $\sim10^5$ trials
we did not see any cases in which the orbital metric had dramatically 
underestimated the mismatch, i.e. where $\relError(m_\F,m_\orb)\sim 2$
in Fig.~\ref{fig:quality_mOrb}. 

It is interesting to note that while the phase metric seems virtually
identical to the orbital metric for ``almost all'' directions in
parameter space, its determinant (and eigenvalues) agree much better
with the (average) $\F$-metric. The small effect of the spin motion of
the earth is negligible for most directions in parameter space, except
for the most degenerate one, where it substantially reduces the degeneracy.

\subsection{The multi-detector $\F$-metric}
\label{sec:f-mismatch-m_f}

As discussed in Sect.~\ref{sec:depend-numb-detect}, the
parameter-space \emph{resolution} of the multi-detector $\F$-metric does
not scale with the number of detectors. Instead, the effect of
combining detectors coherently results in a noise-weighted average of
contributions from different detectors. This averaging operation
\eqref{eq:69} would be expected to decrease the effects of the
antenna-pattern functions $a^\X(t)$ and $b^\X(t)$, as well as the
detector-specific Doppler modulation $\Delta\phi^\X(t)$ of the signal
phase \eqref{eq:77}.  
\begin{figure*}[h!tbp]
  \centering
  \mbox{
    \parbox{0.5\textwidth}{
      \raggedright (a)\\
      \includegraphics[width=0.5\textwidth,clip]{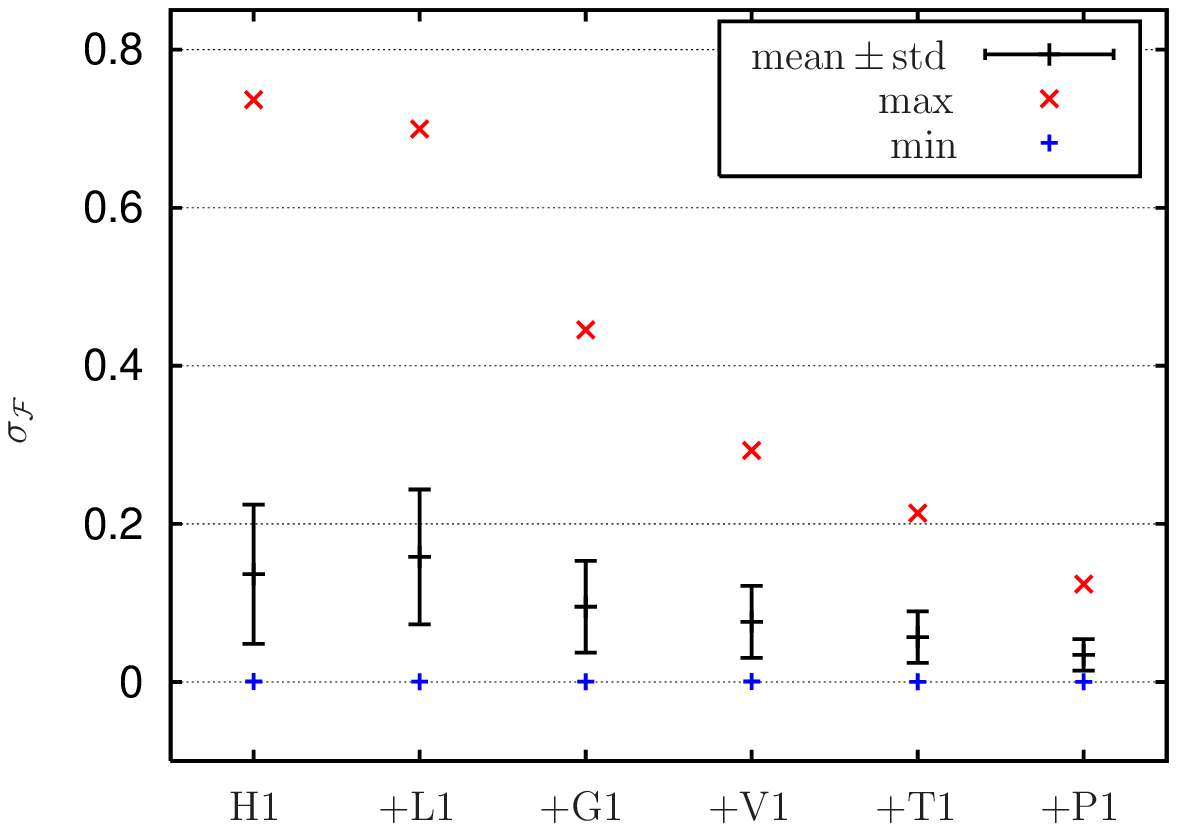}}
    \parbox{0.5\textwidth}{
      \raggedright (b)\\
      \includegraphics[width=0.5\textwidth,clip]{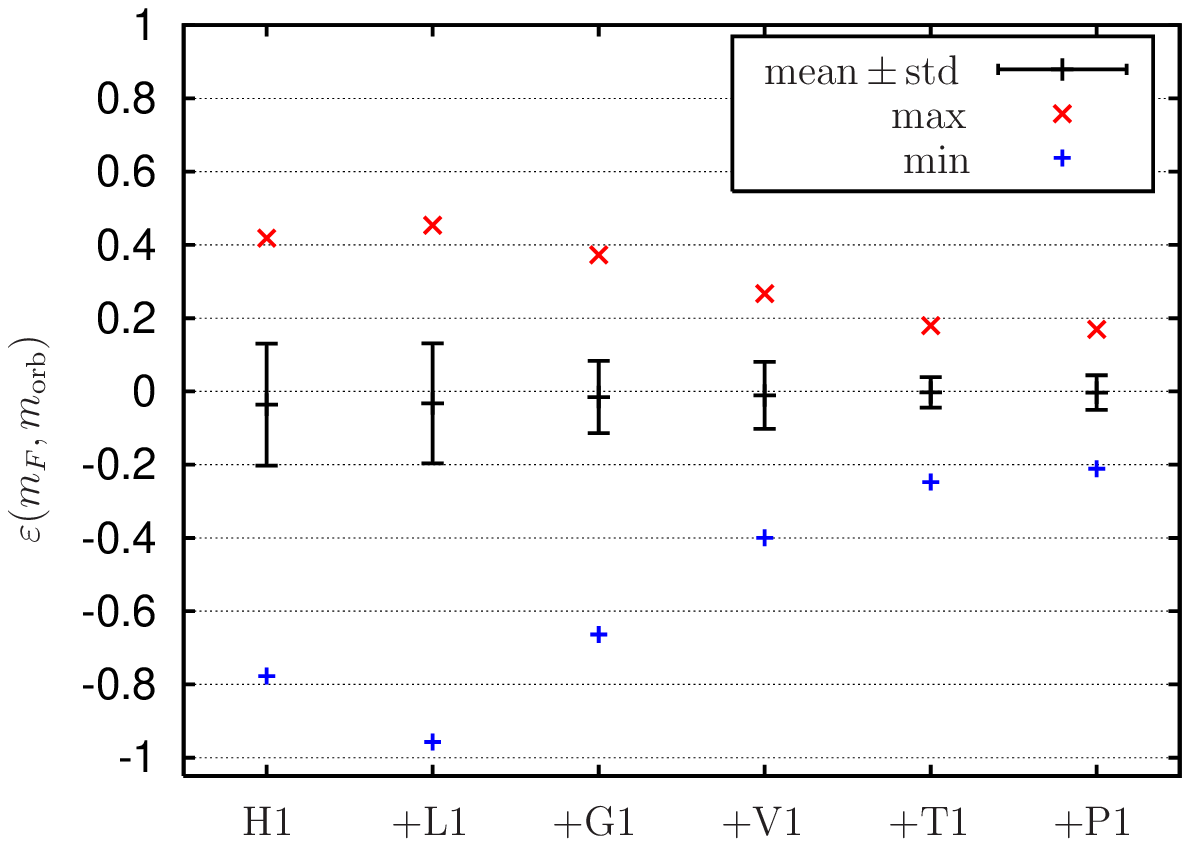}}
  }
  \caption{Distribution of (a) intrinsic relative $\F$-metric
    uncertainty $\uncertF$ and 
    (b) relative errors $\relError(m_\F,m_\orb)$ as functions of
    the coherent combination of (assumed \emph{equal-noise})
    detectors, for $T=55$~hours. 
    Plotted are the mean, standard deviation and extremal values of the  
    distribution for each detector combination. The assumed detectors
    (location and orientation only) are: 'H1' = LIGO Hanford, 'L1' =
    LIGO Livingston, 'G1' = GEO600, 'V1' = Virgo, 'T1' = TAMA, 'P1' =
    Caltech 40m.}    
  \label{fig:multiIFO_convergence}
\end{figure*}
In Fig.~\ref{fig:multiIFO_convergence} we see indeed that both the intrinsic
uncertainty $\uncertF$ of the $\F$-metric family, as well as its
relative difference to the orbital metric decrease with the number of
detectors. 
These results are based on a Monte-Carlo simulation with $\sim40,000$
randomly chosen parameters (see Sect.~\ref{sec:monte-carlo-expl}), for
a fixed observation time of $T=55$~hours, and using between one and six
coherently combined detectors.  For the sake of this example, we have
made the (obviously unrealistic) assumption that all 6 detectors
have the same noise floor, otherwise the convergence would be much
weaker. 
Note that we would find exactly the same mismatch convergence for the phase
metric as that shown in Fig.~\ref{fig:multiIFO_convergence}(b) for the
orbital metric.

\subsection{Comparing  metric predictions to measured mismatches} 
\label{sec:comp-meas-snr}

In order to validate the theoretical $\F$-statistic mismatch $m_\F$,
derived in Sect.~\ref{sec:neutr-star-spec}, we compare it to the
measured relative SNR loss $m_0$ in a simulated mismatched search. 
This is done by first generating (using
\texttt{lalapps\_Makefakedata}~\cite{lalapps}) a fake signal with
parameters $\{\vA, \vDoppler_\sig\}$ picked at random (see
Sect.~\ref{sec:monte-carlo-expl}).  
We then measure (using \texttt{lalapps\_ComputeFStatistic}~\cite{lalapps}) 
the perfectly-matched SNR $\rho(0)$ at the signal location
$\vDoppler_\sig$, and the mismatched SNR $\rho(\Delta\vDoppler)$ at
an offset Doppler position $\vDoppler_\sig + \Delta\vDoppler$. 
The offsets $\Delta\vDoppler$ were picked at random using the
algorithm described in Sect.~\ref{sec:monte-carlo-expl}. 
The \emph{measured} $\F$-mismatch $m_0$ is then given simply by the
definition Eq.~\eqref{eq:53}.
The comparison of the measured mismatches $m_0$ to the theoretical
approximation reveals a problem with the metric approximation for
large angular offsets (in natural units), in particular for $\dOm \gtrsim 5$.  
The origin of this ``metric failure'' can be understood in terms of
the metric curvature and will be discussed in more detail in
\ref{sec:metric-problems-dom}. 
In the meantime we simply remove this known source of errors by
excluding points with $\dOm>5$, which affects less than $1\%$ of the
Monte-Carlo trials. 
\begin{figure*}[h!tbp]
  \centering
  \mbox{
    \parbox{0.5\textwidth}{
      \raggedright (a)\\
      \includegraphics[width=0.5\textwidth,clip]{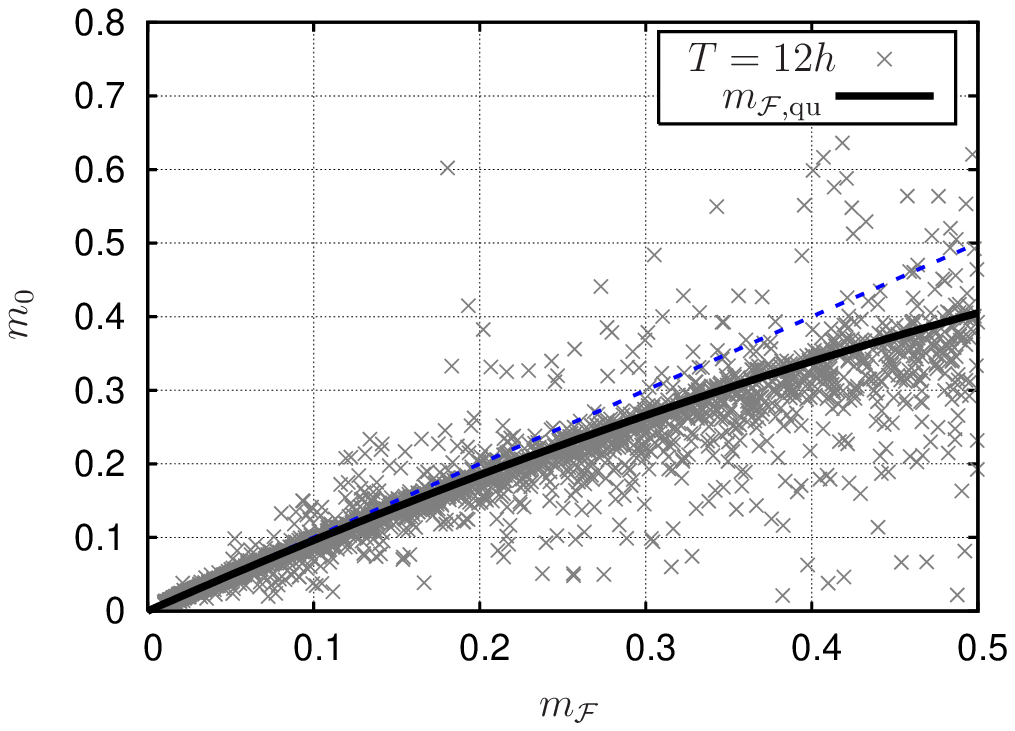}}
    \parbox{0.5\textwidth}{
      \raggedright (b)\\
      \includegraphics[width=0.5\textwidth,clip]{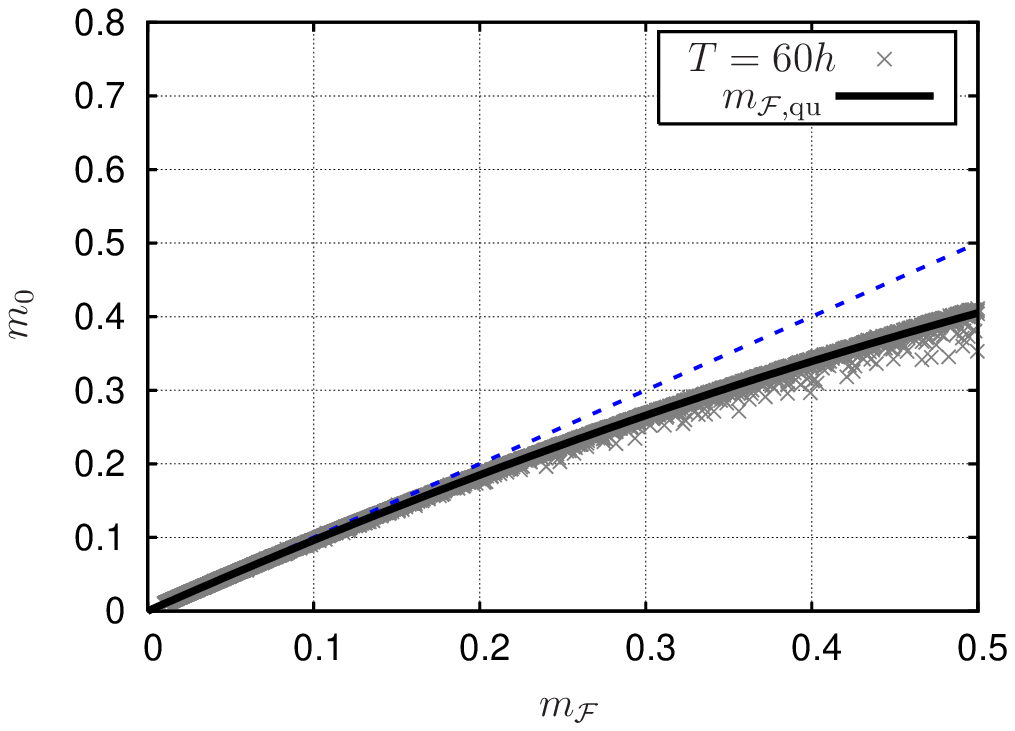}}}
  \caption{Measured mismatches $m_0$ versus predicted $\F$-mismatches
    $m_\F$ for observations times of (a) $T=12$~hours and (b)
    $T=60$~hours, omitting points with $\dOm>5$. 
    The dashed line indicates the identity function.}
  \label{fig:m0_mF}
\end{figure*}
Fig.~\ref{fig:m0_mF} shows the measured mismatches $m_0$ as a function
of the prediction $m_\F$, for observation times of (a) $T=12$~hours and
(b) $T=60$~hours.
We see that there is a substantial scattering of measured
mismatches in the case of $T=12$~hours, which has virtually
disappeared for $T=60$~hours. 
This second type of ``metric failure'', which only affects short
observation times, will be discussed in \ref{sec:metr-probl-tless}. 

Another effect seen in these figures is a systematic deviation of
$m_0$ with respect to $m_\F$ with increasing mismatches, which becomes
noticeable at around $m_\F \gtrsim 0.15$.  
This deviation would be suspected due to higher-order corrections
$\O(\Delta\Doppler^3)$ with respect to the local metric expansion,
and is found to be roughly independent of the observation time $T$. 
We can approximate this systematic deviation by an empirical quadratic
correction of the form     
\begin{equation}
  \label{eq:118}
  m_{\F,\qu}(m_\F) = m_\F  - 0.38 \,m_\F^2 \,, 
\end{equation}
which can be used in comparisons to measured mismatches, as it
compensates for the systematic ``drift'' in $m_0$. 
Note that this deviation makes the metric a conservative
over-estimate, as the actual mismatches tend to be smaller than the
predicted ones.

\subsubsection{Metric problems for $T\lesssim 1$~day}
\label{sec:metr-probl-tless}

The strong scattering of mismatches seen in Fig.~\ref{fig:m0_mF}(a) for
short observation times can be attributed to an intrinsic property of
the $\F$-statistic: namely, the ``local'' parameter-space structure for
short observation times is not very well approximated by a quadratic
form in offsets. 
Note that obviously the metric approximation is valid in the strict
\emph{local} sense of sufficiently small offsets, but in
practice we are more interested in a finite ``local'' region of small
mismatches, $m \lesssim 0.5$ say.
\begin{figure}[h!tbp]
  \includegraphics[width=0.5\textwidth,clip]{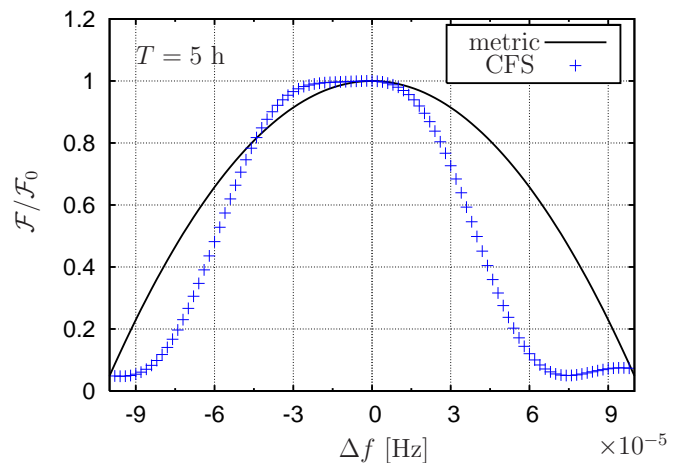}
  \caption{Comparison of the measured decrease in $\F$-statistic (CFS)
    as a function of frequency offset $\Delta f$, and the
    corresponding metric prediction $1 - m_\F$, for an observation time
    of $T=5$~hours.
    [Parameters: $f=100$~Hz, $\cos\iota= 0.95$, $\psi = 2.25$,
    $\alpha=5.98$, $\delta=0.09$, $\fdot=0$] 
  }   
  \label{fig:badMetric}
\end{figure}
\begin{figure}[h!tbp]
  \includegraphics[width=0.5\textwidth,clip]{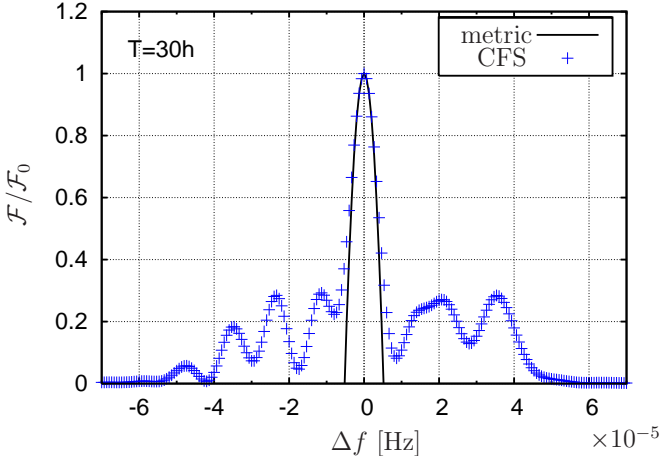}
  \caption{Same as Fig.~\ref{fig:badMetric} for an observation time
    of $T=30$~hours.
  }   
  \label{fig:badMetric30}
\end{figure}
An illustrative example for this is shown in
Figs.~\ref{fig:badMetric}, \ref{fig:badMetric30},  
where we have measured the $\F$-statistic (using
\texttt{lalapps\_ComputeFStatistic}) over a $\sim 10^{-4}$~Hz
band around the true signal frequency, with the other Doppler
parameters ($\alpha,\delta,\fdot$) held fixed at their correct values.
For an observation time of $T=5$~hours, the quadratic decrease predicted
by the $\F$-metric is clearly not a good approximation, as can be seen in
Fig.~\ref{fig:badMetric}: depending on the sign of the frequency
offset, the metric would either substantially under- or
overestimate the mismatch. This effect decreases rapidly with
observation time, and for $T=30$~hours (and the same signal), the 
decrease of the $\F$-statistic is ``locally'' much better approximated by
the metric, as seen in Fig.~\ref{fig:badMetric30}.
This effect can be understood as follows: the quadratic decrease of
the $\F$-statistic predicted by the metric has a typical width in
frequency of $\Delta f_0 \sim 1/T$ (cf. Sect.~\ref{sec:natur-units-doppl}). 
However, the spin motion of the earth creates ``side-lobes'' in
frequency space at offsets of a few
$f_\spin=1/$day$\sim10^{-5}$~Hz, which are clearly visible in  
Fig.~\ref{fig:badMetric30}.   
For observation times much smaller than a day, these side-lobes
are therefore not well separated from the main peak and will
substantially alter its quadratic form, e.g. by creating ``plateaus'' 
and steeper ``cliffs'', as seen in Fig.~\ref{fig:badMetric}. 
As the observation time gets longer, i.e. $T\gtrsim1$~day, the main
peak becomes well separated, and therefore well described by the
metric approximation.

\subsubsection{Metric problems for $\dOm > 5$}
\label{sec:metric-problems-dom}

\begin{figure}[h!tbp]
  \includegraphics[width=0.5\textwidth,clip]{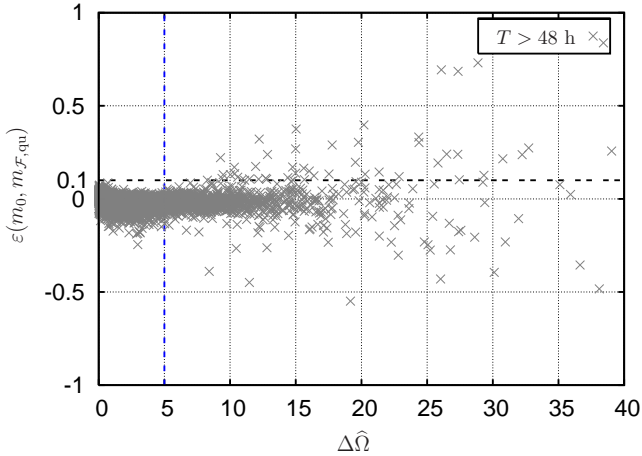}
  \caption{Relative errors $\relError(m_0,m_{\F,\qu})$ versus
    angular offset $\dOm$ in natural units, for observation times 
    $T > 48$~hours. 
    For better readability of the figure, the displayed range of
    angular offsets is limited to $\dOm\le 40$.
}  
  \label{fig:relError_dOm}
\end{figure}

As already mentioned, another problem with the metric approximation
affects points with large angular offsets $\dOm$, and is present even
for long observation times.  
This can be seen in Fig.~\ref{fig:relError_dOm}, showing the relative
errors $\relError(m_0,m_{\F,\qu})$ as a function of angular offset
$\dOm$, for observation times $T>48$~hours.  
For angular offsets $\dOm<5$, the relative errors stay below
$\relError(m_0, m_{\F,\qu})<0.1$, but with increasing $\dOm$, the
errors start to spread out substantially.   

The reason for this metric failure can be traced to
the curvature of the metric on the sky. While the metric 
ellipses have constant orientation as functions of frequency and
spindown, their orientation changes with sky position. 
This curvature is closely related to the global ``circles in the sky'' (CiS)
structure discussed in \cite{prix05:_circles_sky}, as the metric
ellipses on the sky are ``tangential'' to these circles.
To first order, the CiS are described by the equation 
\begin{equation}
  \label{eq:5}
  f\left( 1 + \vn\cdot \vec{V}/c\right) = f_\sig\left(1 + \vn_\sig\cdot\vec{V}/c\right)\,,
\end{equation}
where $f$ and $\vn$ are the ``target'' frequency and sky position
respectively, while $f_\sig$ and $\vn_\sig$ are the corresponding
signal parameters. 
\begin{figure}[h!tbp]
  \includegraphics[width=0.5\textwidth,clip]{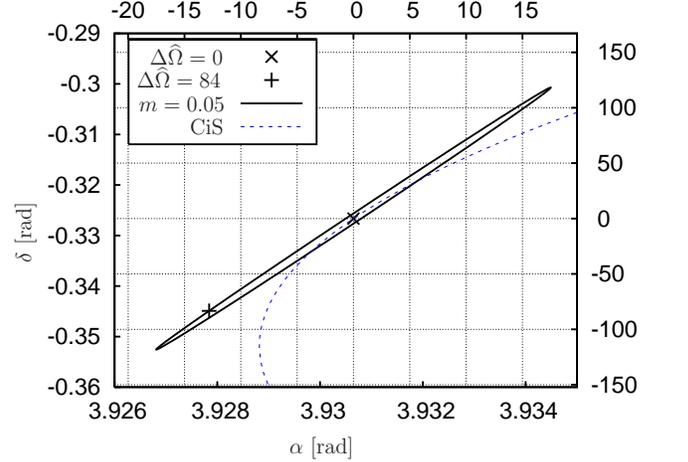}
  \caption{Metric ``failure'' for large angular offset $\dOm=84$ at
    $T=68$~hours. 
    The point 'x' indicates the signal location, '+' shows an offset
    location on the $m=0.05$ iso-mismatch ellipse with a measured
    mismatch of $m_0\sim1$.  
    The dashed line indicates the ``circle in the sky'' \eqref{eq:5}
    defined by the signal location.
    The top and right-hand axes show the offsets in natural units. 
    [Parameters: $f= 186.34$~Hz, $\alpha = 3.93$, $\delta = -0.33$,
    $\fdot =  1.7\times10^{-10} s^{-2}$, $\cos\iota = 0.4$, $\psi = 1.65$   
    ]
  }
  \label{fig:badMetricHighT}
\end{figure}
The problem stems from the curvature of the $\F$-statistic circles in
the sky-coordinates $\alpha$, $\delta$, which is why the CiS is only
locally well approximated by the respective mismatch ellipses. 
This effect is seen clearly in Fig.~\ref{fig:badMetricHighT},
which shows an extreme example of such a metric ``failure'' due to
large $\dOm$. In this figure, '+' indicates a sky position on the
$m_\F=0.05$ iso-mismatch ellipse, which has an angular offset from the
signal of $\dOm=84$. Contrary to the predicted mismatch, the measured
mismatch at this point is $m_0 \sim 1$, as the $\F$-statistic
decreases rapidly away from the CiS (indicated by the dashed line).

\subsubsection{Summary of metric validation}
\label{sec:summ-metr-valid}

\begin{figure}[h!tbp]
  \centering
  \mbox{
    \includegraphics[width=0.5\textwidth,clip]{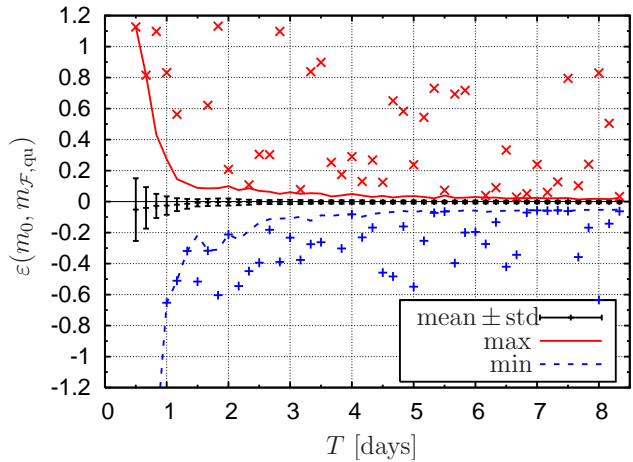}
  }
  \caption{Distribution of relative errors $\relError(m_0,m_{\F,\qu})$
    as a function of observation time $T$.
    Plotted are the mean, standard deviation and extremal values of the  
    distribution for each value of $T$, excluding points with $\dOm>5$.
    The points 'x' and '+' indicate the respective maximum and minimum
    errors when \emph{including} results with $\dOm>5$.
  } 
  \label{fig:err_mFqu_m0}
\end{figure}

Fig.~\ref{fig:err_mFqu_m0} summarizes the relative errors 
$\relError(m_0,m_{\F,\qu})$ between the (quadratically corrected)
predictions \eqref{eq:118} and the measured mismatches $m_0$, as a
function of observation time $T$.  
We see that, as expected, excluding large angular
offsets ($\dOm>5$) yields a substantially improved agreement between
the metric prediction and the measurements, as it eliminates the type
of ``metric failures'' discussed in \ref{sec:metric-problems-dom}. 
We also see that the relative errors can still be quite large for
short observation times $T\lesssim 1$~day, as discussed in
\ref{sec:metr-probl-tless}, and that these errors decrease rapidly
with $T$. For observation times of $T\gtrsim1$~day, the average error
is below a few percent. We can therefore conclude that the agreement
of the $\F$-metric with the measurements is very good in the domain of
applicability of the metric approximation. 

\section{Discussion}
\label{sec:conclusions}

We have derived a formalism for the general parameter-space metric of
the multi-detector $\F$-statistic, and we have explicitly computed the
metric for signals from isolated spinning neutron stars.   
We have shown that there exists a family of $\F$-metrics, parametrized
by the two (unknown) amplitude parameters $\psi$ and $\cos\iota$.
We explicitly derived the extremal ``mismatch bounds'' 
(i.e. the maximum and minimum possible mismatches) of the
$\F$-metric family, and we introduced an average $\F$-metric, which is
independent of the unknown amplitude parameters.
We have shown that the multi-detector $\F$-metric does not scale with
the number of detectors. Combining detectors coherently therefore does
not increase the required number of templates. 
In the long-duration limit ($T\gtrsim 1$~month), we found that the
$\F$-metric family converges towards a simple orbital metric
$g^\orb_{ij}$, which neglects both the amplitude modulation and the
phase modulation caused by the diurnal rotation of the Earth. 

Both the orbital and the closely related phase metric provide
relatively good mismatch approximations in practice, and the quality
of these approximations improves with longer observation times and
with the number of coherently-combined detectors.

The orbital metric, however, while virtually identical to the phase
metric for almost all directions in parameter space, was found to be
substantially more degenerate for observation times shorter than a
month. This has important consequences for the covering problem of the
parameter space and requires further study. Finally, we have
identified two regimes in which the local metric approximation itself
is not very reliable: namely, for observation times $T\lesssim 1$~day,
and for large angular offsets $\dOm \gtrsim 5$ (in natural units). 

\appendix

\section{Alternative derivation of the $\F$-metric}
\label{sec:altern-proj-onto}

A more elegant derivation of the $\F$-metric \eqref{eq:34} can be
obtained~\cite{cutler_private} by projecting the full parameter-space
metric in $\{\vA,\vDoppler\}$ into the reduced parameter space
$\vDoppler$ of the $\F$-statistic.
We denote the full parameter space as $\parm = \{\vA,\vDoppler\}$, and
we use index conventions $\parm^a = \{\A^\mu, \Doppler^i\}$.
Recall the form of the log-likelihood ratio \eqref{eq:26}, namely
\begin{equation}
  \label{eq:3}
  \ln\Lambda(\detVec{x};\parm) \equiv \left(\detVec{x}|\detVec{s}(\parm)\right) 
  - {1\over2}\left(\detVec{s}(\parm)|\detVec{s}(\parm)\right)\,.
\end{equation}
If the data contains a signal with parameters $\parm_\sig$, 
i.e. \mbox{$\detVec{x}(t) = \detVec{n}(t) + \detVec{s}(t; \parm_\sig)$}, 
and if the target position $\parm$ is ``close'' to the signal location, 
i.e. $\parm = \parm_\sig + \Delta\parm$  for ``small'' $\Delta\parm$,
then the  expectation value of $\ln\Lambda$ can be expanded as 
\begin{equation}
  \label{eq:44}
  2 E\left[ \ln\Lambda(\parm| \parm_\sig) \right]
  = \left( \detVec{s} | \detVec{s} \right)
  - \left( \partial_a \detVec{s} | \partial_b \detVec{s} \right)\Delta\parm^a \Delta\parm^b 
  + \O(3) \,,
\end{equation}
so the full parameter-space metric $\gt_{a b}$ is found as
\begin{equation}
  \label{eq:45}
  \gt_{a b} =  { (\partial_a \detVec{s} | \partial_b \detVec{s}) \over
    (\detVec{s}|\detVec{s})}\,. 
\end{equation}
This expression is sometimes referred to as the \emph{normalized} Fisher
matrix \cite{krolak04:_optim_lisa}. As mentioned in the introduction,
this ``canonical'' metric differs from a definition often found in the literature 
(e.g.~\cite{owen96:_search_templates,bala96:_gravit_binaries_metric,cornish05:_detec_lisa}),
which is based on a somewhat more \emph{ad-hoc} measure of the ``match'',
namely $M \equiv\left(s(\parm_\sig)|s(\parm)\right)$, 
instead of the full log-likelihood \eqref{eq:3}.
As it turns out, both definitions result in the same phase metric
\eqref{eq:122} when considering constant-amplitude signals (after
minimizing the mismatch \eqref{eq:45} over the unknown amplitude). 
In general, however, the canonical definition \eqref{eq:45}, and
correspondingly \eqref{eq:53}, is more directly relevant to the
covering problem, as it describes the relative loss of detection
statistic.   
For the assumed general form of the signal \eqref{eq:133}, we have
\begin{equation}
  \label{eq:97}
  \rho^2(0) \equiv (\detVec{s}|\detVec{s}) = \A^\mu \M_{\mu\nu} \A^\nu\,,
\end{equation}
and the respective derivatives with respect to the amplitude and Doppler
subspaces are given by 
\begin{eqnarray}
  \partial_\mu \detVec{s}  &=& {\partial \detVec{s} \over \partial
    A^\mu} = \detVec{h}_\mu(t;\vDoppler)\,,  \label{eq:46}\\ 
  \partial_i \detVec{s} &=& {\partial \detVec{s} \over \partial
    \Doppler^i} = \A^\mu \, \partial_i \detVec{h}_\mu\,.
  \label{eq:47} 
\end{eqnarray}
The full metric $\gt_{a b}$ therefore consists of the three blocks
(with respect to the two subspaces): 
\begin{equation}
  \label{eq:4}
  \gt_{a b} = \rho^{-2}(0)\, \left(
    \begin{array}{c c}
      \M_{\mu\nu} & \A^\nu \R_{\mu\nu i} \\[\vsep]
      \A^\nu \R_{\mu\nu i} & \A^\alpha h_{\alpha\beta i j}\A^\beta\\
    \end{array}
  \right)\,,
\end{equation}
in terms of $h_{ij}$ and $\R_i$ defined in \eqref{eq:28} and
\eqref{eq:21}. 
The $\F$-metric $g^\F_{ij}$ in the Doppler subspace can be regarded as
the distance corresponding to given Doppler offsets 
$\dth^i = \Delta\Doppler^i$, \emph{minimized} over the amplitude
offsets $\dth^\mu = \Delta\A^\mu$, i.e.
\begin{eqnarray}
  \label{eq:128}
  g^\F_{ij} \,\Delta\Doppler^i\,\Delta\Doppler^j &\equiv&
  \min_{\Delta\!\A^\mu}\,\gt_{a b}\,\dth^a\,\dth^b\,.
\end{eqnarray}
This can be minimized trivially, since it is a quadratic function in
$\Delta\A^\mu$, and the ``compensating'' amplitude mismatches are
\begin{equation}
  \label{eq:138}
  \dth^\mu = -\gh^{\mu\nu}\,\gt_{\nu i}\, \dth^i\,,
\end{equation}
where $\gh^{\mu\nu}$ is the inverse matrix of $\gt_{\mu\nu}$, i.e.
$\gh^{\mu\alpha}\gt_{\alpha\nu} = \delta^\mu_\nu$. Inserting this into
\eqref{eq:128} we obtain
\begin{equation}
  \label{eq:139}
  g^\F_{ij} = \gt_{ij} - \gt_{i\mu}\gh^{\mu\nu}\gt_{\nu j}\,,
\end{equation}
which corresponds to the \emph{projection} of the full metric 
$\gt_{a b}$ into the Doppler subspace. Using the explicit components
\eqref{eq:4}, we find
\begin{eqnarray}
  \label{eq:52}
  g_{i j}^\F(\vA) &=& \rho^{-2}(0)\, \vA\left[ 
    h_{i j} - \R\trans_i \M^{-1} \R_j \right]\vA \\
  &=& { \vA \cdot \G_{i j} \cdot \vA \over \vA\cdot\M\cdot\vA}\,,
\end{eqnarray}
in perfect agreement with the earlier result \eqref{eq:130}, which was
obtained in a more straightforward, but somewhat more tedious
calculation.

\section{$\F$-metric for low-frequency signals}
\label{sec:keep-ampl-funct}

In the case of low frequencies $f_\sig$ (which would be relevant for LISA)
and/or short observation times $T$, where $f_\sig\,T\not\gg 10^4$, 
we cannot use the simplifying approximation of
Sect.~\ref{sec:f-metric-ground}.
We can nevertheless proceed in the same way: using the expansion
\eqref{eq:40} and keeping only leading-order terms, we can express
\eqref{eq:28} as 
\begin{eqnarray}
  h_{\mu\nu\,i j} &\approx&
  {1\over 2} \Sinv T \,
  \left( \begin{array}{c c c c}
      P^1_{i j} &  P^3_{i j} &  0   &  P^4_{i j} \\
      P^3_{i j} &  P^2_{i j} & -P^4_{i j} &  0   \\
      0   & -P^4_{i j} &  P^1_{i j} &  P^3_{i j} \\
      P^4_{i j} &  0   &  P^3_{i j} &  P^2_{i j} \\
    \end{array}\right)\,,
\end{eqnarray}
in terms of the four independent components
\begin{equation}
  \begin{array}{rcl}
    P^1_{i j} &=& \avS{\partial_i a\, \partial_j a}
    + \avS{a^2\,\partial_i\phi\,\partial_j\phi}\,,\\[\vsep]
    P^2_{i j} &=& \avS{\partial_i b\, \partial_j b} + 
    \avS{b^2 \,\partial_i\phi\, \partial_j\phi}\,,\\[\vsep]
    P^3_{i j} &=& \avS{\partial_i a\, \partial_j b}
    + \avS{a\,b\,\partial_i\phi\,\partial_j\phi}\,,\\[\vsep]
    P^4_{i j} &=& \avS{b\,\partial_i a \, \partial_j \phi}
    - \avS{a\, \partial_i b\, \partial_j\phi}\,,\\
  \end{array}
\end{equation}
with implicit symmetrization in $i,j$.
In a similar manner we calculate $\R_{\mu\nu\,i}$,
defined in \eqref{eq:21}, which yields
\begin{equation}
  \label{eq:59}
  \R_{\mu\nu\,i} \approx {1\over2}\Sinv T\, 
  \left( \begin{array}{c c} 
      \widehat{\R}_i  &  \widetilde{\R}_i \\
      - \widetilde{\R}_i  &  \widehat{\R}_i \\
    \end{array} \right)\,,
\end{equation}
where the 2x2 matrices $\widehat{\R}_i$ and $\widetilde{\R}_i$ are
defined as 
\begin{equation}
  \widehat{\R}_i \equiv \left(\begin{array}{c c}
      R^{11}_i & R^{12}_i \\
      R^{21}_i & R^{22}_i \\
    \end{array}\right)\,,
  \;\textrm{and}\;
  \widetilde{\R}_i \equiv \left(\begin{array}{c c}
      R^{13}_i & R^{14}_i \\
      R^{14}_i & R^{24}_i \\
    \end{array}\right)\,,
\end{equation}
in terms of
\begin{equation}
  \begin{array}{c}
  R_i^{11} = \avS{a\,\partial_i a}\,,\quad
  R_i^{12} = \avS{a\,\partial_i b}\,,\\[\vsep]
  R_i^{21} = \avS{b\,\partial_i a}\,,\quad
  R_i^{22} = \avS{b\,\partial_i b}\,,\\[\vsep]
  R_i^{13} = \avS{a^2\partial_i \phi}\,,\quad
  R_i^{24} = \avS{b^2\partial_i \phi}\,,\\[\vsep]
  R_i^{14} = \avS{a\,b\,\partial_i \phi}\,.\\
  \end{array}
\end{equation}
Note that $\widehat{R}_i$ only contains derivatives of the
antenna-pattern functions $a^\X$, $b^\X$ (which where neglected in
Sect.~\ref{sec:f-metric-ground}), while $\widetilde{R}_i$ only
contains derivatives of the phase. 
As a consequence of this block structure, one finds
\begin{equation}
  \left\{\trans\R_i\M^{-1}\R_j\right\}_{\mu\nu} \approx {1\over 2}\Sinv T
  \left( \begin{array}{c c c c}
      Q_{i j}^1 & Q_{i j}^3  &   0   & Q_{i j}^4 \\
      Q_{i j}^3 & Q_{i j}^2  & -Q_{i j}^4  &  0  \\
      0   & -Q_{i j}^4 &  Q_{i j}^1  & Q_{i j}^3 \\
      Q_{i j}^4 &  0   &  Q_{i j}^3  & Q_{i j}^2 \\
    \end{array}\right)\,,
\end{equation}
with the four independent components 
\begin{equation}
  \begin{array}{rcl}
    D\, Q^1_{i j} &=& A\left[ R^{21}_i R^{21}_j  + R^{14}_i R^{14}_j \right]
    + B\left[ R^{11}_i R^{11}_j + R^{13}_i R^{13}_j \right] \\[\vsep]
    && - 2 C \left[ R^{11}_i R^{21}_j + R^{13}_i R^{14}_j \right]\,,\\[\vsep]
    D\, Q^2_{i j} &=& A\left[ R^{24}_i R^{24}_j + R^{22}_i R^{22}_j \right]
    + B\left[ R^{14}_i R^{14}_j + R^{12}_i R^{12}_j \right] \\[\vsep]
    && - 2C\left[ R^{14}_i R^{24}_j + R^{12}_i R^{22}_j \right]\,,\\[\vsep]
    D\, Q^3_{i j} &=& A\left[ R^{14}_i R^{24}_j + R^{22}_i R^{21}_j \right]
    + B\left[ R^{14}_i R^{13}_j + R^{12}_i R^{11}_j \right] \\[\vsep]
    && \hspace*{-1cm} - C \left[ R^{24}_i R^{13}_j + R^{14}_i R^{14}_j + R^{11}_i R^{22}_j
      + R^{12}_i R^{21}_j \right]\,, \\[\vsep]
    D\, Q^4_{i j} &=& A\left[ R^{21}_i R^{24}_j - R^{14}_i R^{22}_j \right]
    + B\left[ R^{11}_i R^{14}_j - R^{12}_i R^{13}_j \right] \\[\vsep]
    && \hspace*{-1cm} - C\left[ R^{11}_i R^{24}_j - R^{13}_i R^{22}_j  + R^{14}_i R^{21}_j
      - R^{12}_i R^{14}_j \right]\,.
  \end{array}
\end{equation}
Putting all the pieces together, we find $\G_{i j}$, defined in
\eqref{eq:37}, in the form:
\begin{eqnarray}
  \G_{\mu\nu} \approx {1\over 2}\Sinv T \,\left( \begin{array}{c c c c}
      m^1 & m^3  &   0   & m^4 \\
      m^3 & m^2  & -m^4  &  0       \\
        0 & -m^4 &  m^1  & m^3 \\
      m^4 &  0   &  m^3  & m^2 \\
    \end{array}\right)\,,
\end{eqnarray}
in terms of the \emph{four} components ($r=1,2,3,4$): 
\begin{equation}
  m^r \equiv \left( P^r_{i j} - Q^r_{i j}
  \right)\,\Delta\Doppler^i \Delta\Doppler^j\,.
\end{equation}
As discussed in Sect.~\ref{sec:f-metric}, the extrema $\mhat_\F$ are
the eigenvalues of $\M^{-1}\cdot\G$, which are the solutions of  
\begin{eqnarray}
  0 &=& \det\M^{-1}\, \det\left[ \G - \mhat_\F\,\M \right]\,,
\end{eqnarray}
or equivalently
\begin{eqnarray}
  0 &=& (m^1 - \mhat_\F A)(m^2 - \mhat_\F B) \nonumber\\
  && \hspace*{4em}- (m^3 - \mhat_\F C)^2 - (m^4)^2\,.
\end{eqnarray}
We see that there are again maximally \emph{two} independent
eigenvalues, namely  
\begin{equation}
  \mhat_\F^{\max|\min}(\vDoppler;\Delta\vDoppler) = \mbar_\F \pm \sqrt{ \mbar_\F^2 - \mtilde^2 }\,,
\end{equation}
with
\begin{equation}
  \label{eq:132}
  \begin{array}{rcl}
    \mbar_\F &=&  (2D)^{-1}\left[B m^1 + A m^2 -2C m^3\right]\,,\\[\vsep]
    \mtilde &=& D^{-1}\left[ m^1 m^2 - (m^3)^2 - (m^4)^2 \right]\,,\\
  \end{array}
\end{equation}
which is formally very similar to the earlier result \eqref{eq:68}.

\begin{acknowledgments}
I am very grateful to Curt Cutler for numerous helpful suggestions,
and I thank Chris Messenger, Andrzej Kr\'olak, Badri Krishnan and Ben 
Owen for useful discussions and comments.
\end{acknowledgments}

\bibliography{biblio}

\end{document}